\documentclass[a4paper,11pt]{article}
\pdfoutput=1 

\usepackage{jinstpub} 
\usepackage[utf8]{inputenc}
\usepackage{lineno}

\usepackage[capitalise,noabbrev,nameinlink]{cleveref}
\newlength{\abc}
\AtBeginDocument{%
\renewcommand{\ref}[1]{\mbox{\Cref{#1}}}

}

\usepackage{siunitx}
\DeclareSIUnit\MIP{MIP}
\DeclareSIUnit\inch{''}

\usepackage{multirow}

\usepackage{textcomp}
\usepackage{xcolor}
\usepackage{graphicx}
\usepackage{mathtools}
\usepackage{subcaption}
\usepackage[LGRgreek]{mathastext}
\usepackage{xspace}
\usepackage{ptdr-definitions}
\usepackage{heppennames2}

\usepackage[export]{adjustbox}
\newcommand{\ind}[0]{\hspace*{\customindent}}

\newcommand{\parag}{\parag\ind}

\newcommand{\tocite}[1]{\textbf{\textcolor{red}{[\ifthenelse{\equal{#1}{}}{?}{#1}]}}}

\newcommand{\neqcm}{\ensuremath{\mathrm{n_{eq}}/\mathrm{cm}^{2}}\xspace}
\definecolor{ttHcol}{RGB}{67,118,201}
\definecolor{ttbbcol}{RGB}{235,230,10}
\definecolor{ttbcol}{RGB}{205,0,10}
\definecolor{tttbcol}{RGB}{255,153,1}
\definecolor{ttcccol}{RGB}{131,38,10}
\definecolor{ttlfcol}{RGB}{81,142,25}
\definecolor{Singletcol}{RGB}{0,204,204}
\definecolor{Vjetscol}{RGB}{104,140,140}
\definecolor{Dibosoncol}{RGB}{1,25,147}
\definecolor{ttVcol}{RGB}{255,102,101}

\title{\boldmath Timing Performance of the CMS High Granularity Calorimeter Prototype}

\collaboration{CMS HGCAL collaboration}

\author[2]{B.~Acar,}
\author[13]{G.~Adamov,}
\author[37]{C.~Adloff,}
\author[26]{S.~Afanasiev,}
\author[45]{N.~Akchurin,}
\author[2]{B.~Akg\"{u}n,}
\author[4]{F.~Alam Khan,}
\author[25]{M.~Alhusseini,}
\author[5]{J.~Alison,}
\author[19]{A.~Alpana,}
% ...
% ...
\affiliation[2]{Bo\u{g}azi\c{c}i University, \\Bebek 34342, Istanbul, Turkey}
% ...
\affiliation[4]{CERN,\\Espl. des Particules 1, 1211 Geneva 23, Switzerland}
\affiliation[5]{Carnegie Mellon University, \\ 5000 Forbes Ave, Pittsburgh 15213, PA, USA}
% ...
\affiliation[13]{Georgian Technical University, \\ 77 Kostava Str 0175, Tbilisi, Georgia}
% ...
\affiliation[19]{Indian Institute of Science Education and Research, \\ Dr. Homi Bhabha Road 411008, Pune, India}
% ...
\affiliation[25]{The University of Iowa,\\ 203 Van Allen Hall, Iowa City, 52242, Iowa, USA}
\affiliation[26]{International Intergovernmental Organization Joint Institute for Nuclear Research JINR, \\ 6 Joliot-Curie St, Dubna 141980, Moscow, Russia}
% ...
\affiliation[37]{National Central University Taipei (NCU),\\No.300, Jhongda Rd 32001, Jhongli City, Taiwan}
% ...
\affiliation[45]{Texas Tech University,\\ Lubbock 79409, TX, USA}
% ...
\emailAdd{Andre.David@cern.ch}

\abstract{
    This paper describes the experience with the calibration, reconstruction and evaluation of the timing capabilities of the CMS HGCAL prototype in the beam tests in 2018.
    The calibration procedure includes multiple steps and corrections ranging from tens of nanoseconds to a few hundred picoseconds.
    The timing performance is studied using signals from positron beam particles with energies between \qty{20}{\GeV} and \qty{300}{\GeV}.
    The performance is studied as a function of particle energy against an external timing reference as well as standalone by comparing the two different halves of the prototype.
    The timing resolution is found to be \qty{60}{\pico\second} for single-channel measurements and better than \qty{20}{\pico\second} for full showers at the highest energies, setting excellent perspectives for the HGCAL calorimeter performance at the HL-LHC.
} 

\begin{document}
\maketitle
\flushbottom

\section{Introduction}
\label{sec:intro}
%Content: context, goals, reminder on state of the art
%\textcolor{red}{Contributors: Arabella, Artur, Roger}

In recent years there has been a growing interest in precision timing for uses beyond the traditional time-of-flight measurements for particle identification.
In anticipation of the very large number of simultaneous interactions that will occur
in a single bunch crossing (pile up) at the High-Luminosity LHC (HL-LHC), both the CMS and ATLAS Collaborations
are developing specialized detectors that can measure the time of the passage
of a charged particle with a precision of a few \qty{10}{\ps}.
These detectors will allow the separation of different proton-proton interactions within a single bunch crossing, which occurs over an interval of about \qty{350}{\ps}.%corresponding to the length of proton bunches crossing at the HL-LHC.

%%%%% How in the monkeys did this get here?
%&\newline

As part of the detector upgrade for the HL-LHC, the CMS Collaboration
will replace the two endcap calorimeters with new, high-granularity, sampling calorimeters (HGCAL, or CE for Calorimeter Endcap).
The electromagnetic sections of HGCAL will be instrumented entirely with silicon sensors, while 
the hadronic sections will be equipped with silicon sensors in the regions of the calorimeter where the total 
fluence is expected to be above \qty{5e13}{\neqcm},
%integrated radiation dose is expected to be above \qty{200}{\kilo Rad},
and with plastic scintillators read out by silicon photomultipliers (SiPMs) elsewhere.
Full details of the CMS endcap calorimeter upgrade may be found in Ref.~\cite{hgcal-tdr:2018}.

In the development program for the HGCAL, a series of tests with single components and prototypes
of the calorimeter have been conducted using early versions of the readout electronics~\cite{HGCTB:JINST2018}.
We report here on the results obtained on the time development of electromagnetic showers
based on data collected with a prototype calorimeter in the H2 beam line of the CERN Super Proton Synchrotron (SPS) in 2018.
Although the prototype did not fully represent the final detector and electronics components, it allowed the main performance features of the final system to be studied. 
In 2016, in a beam test with single prototype silicon diode sensors at the SPS the intrinsic timing resolution of the silicon sensors was evaluated~\cite{HGCTB:JINST2018}. 
In the beam test reported here, the timing performance of a 28-layer electromagnetic calorimeter with a full analog read-out chain with precision timing in each channel was measured.

Previous publications have described the construction and the data acquisition system of this prototype calorimeter~\cite{beamtest-daq:2018,beamtest-calibration:2018}, and the energy resolution, linearity, and position and angular resolutions for both positrons and charged pions~\cite{beamtest-electrons:2018,beamtest-pions:2018}.

%In this report we present the results of timing measurements obtained with electromagnetic showers. 

%\textbf{Merge with below!}
%
%The largest prototype of the HGCAL up-to-date was assembled for the SPS beam test in October 2018. 
%Although it did not yet feature the final detector and electronics components, it allowed to study the main performance features of the final system. 
%A previous beam test of single prototype silicon diode sensors has been performed in 2016 at SPS in order to evaluate the intrinsic timing resolution of the silicon sensor~\cite{hgcal-tb2016}.
%In contrast to that test, this beam test of full silicon module prototypes allowed for the characterization of the full analog read-out chain of the timing measurement.
\section{Experimental setup}
\label{sec:expSetup}

\subsection{Infrastructure at the SPS H2 beam line}
%The analyses were performed with the 
Data were collected in October 2018 at the H2 beamline~\cite{h2_beamline} at the CERN-SPS North Area. 
Full details of the beamline may be found in Ref.~\cite{beamtest-electrons:2018}.  This study used a secondary beam of positrons with a momentum of up to \qty{300}{\GeV/c}.% that will be referred to as electrons in the following for simplicity. 

\begin{figure}[tb]
	\centering
	\includegraphics[width=0.99\textwidth]{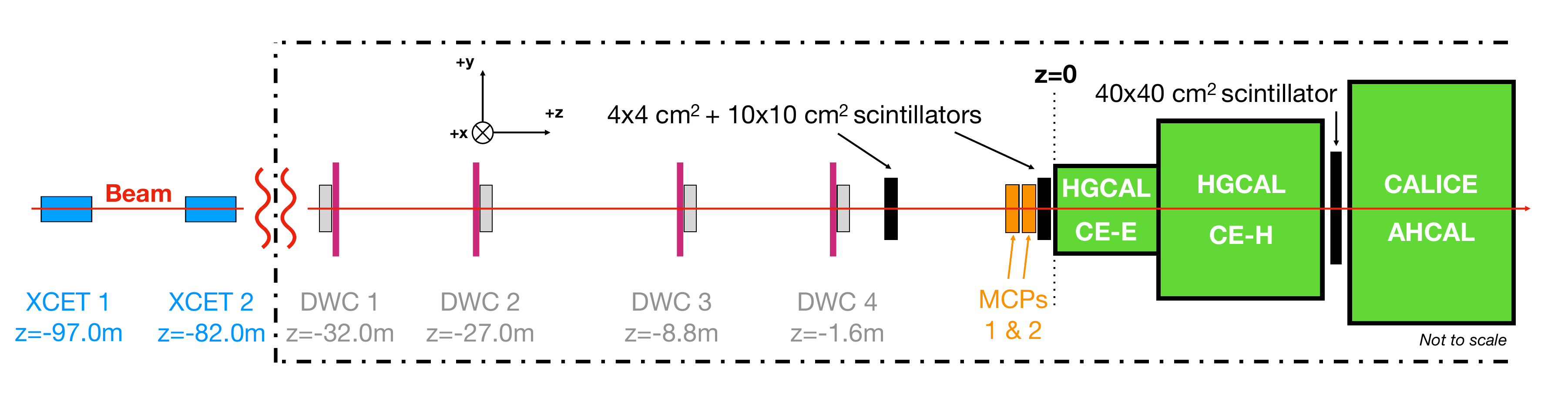}
	\caption{The HGCAL prototype experimental setup used during the 2018 test beam. The electromagnetic section, which is the focus of this paper, is denoted as CE-E. It is preceded (to the left) by multiple ancillary beam instrumentation detectors, and is followed by detector prototypes dedicated to the study of hadron showers.
	\label{fig:example_H2beamline}
	}
\end{figure}

%layout is shown in figure  \ref{fig02:CEE-layout}. 

\begin{figure*}[!ht]
	\centering
	\includegraphics[width=0.49\textwidth]{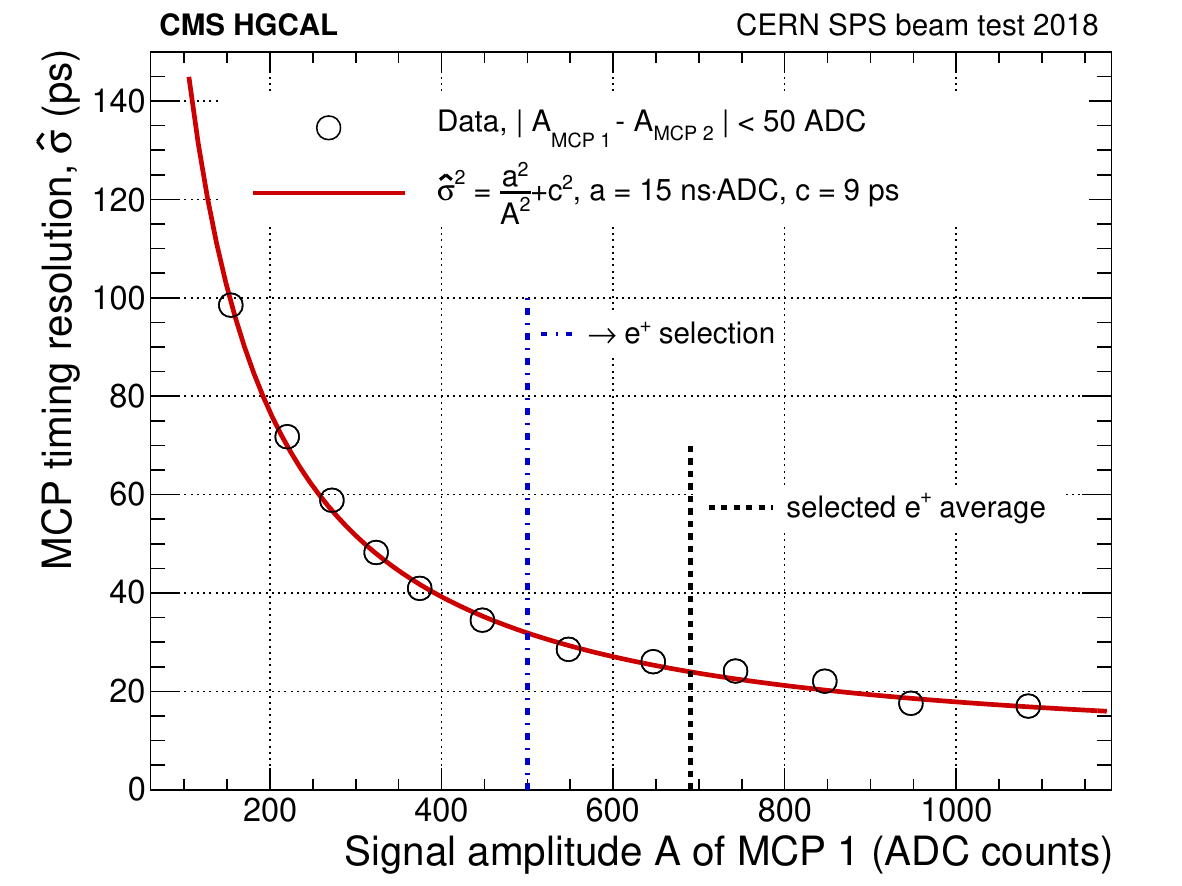}
	\caption{Single MCP time resolution estimated as $\sigma/\sqrt{2}$, where $\sigma$ comes from a Gaussian fit to the distribution of time differences between the two MCPs operated in front of the HGCAL prototype calorimeter.
	Given the identical structure of the two MCP detectors, their timing resolution was assumed to be identical.
	\label{fig:mcpreso}
	}
\end{figure*}

The experimental setup is sketched in \ref{fig:example_H2beamline}. Two micro-channel plate (MCP) detectors~\cite{BRIANZA2015216} were placed immediately upstream of the CE-E prototype to provide a reference measurement of the time-of-arrival of the incident particles.
The sensitive area of the MCP detectors was about \qty{1}{\cm\squared}, defining the transverse extent of the accepted events.
% (cf. Fig. \ref{fig:example_H2beamline}). 
One of the MCP detectors (MCP 1) was used as a timing reference, while the second one was used for cross-calibration to obtain the timing resolution as a function of the signal amplitude, shown in \ref{fig:mcpreso}.
The MCP timing resolution depends on the deposited charge and the asymptotic timing resolution of a single MCP was found to be about \qty{9}{\ps}.
The average resolution for positrons selected for this study was measured to be about \qty{25}{\ps} corresponding to MCP signals of about \qty{700}{ADC~counts}.

Two scintillators, used to generate the event trigger, were placed
before and after the MCP detectors. 
Four delay wire chambers used to determine the impact position of beam particles were placed upstream of the trigger scintillators. 
Further details of the experimental setup can be found in Ref.~\cite{beamtest-electrons:2018}. 

%\begin{figure*}[!ht]
	%  \begin{center}
		%    \includegraphics[width=1.\textwidth, angle=-90]{figures/Chapter3/HGCalTB2018_CE-E.pdf}
		%  \end{center}
	%\caption{ 
		%Layout for the CE-E prototype. This prototype was built of 14 cassettes, where each cassette carries two hexagonal modules. During the beam tests, the beam entered the prototype through the first cassette. 
		%In a hexagonal module, the Si sensor was located between the copper-tungsten plate (dark grey) and the Hexaboard (green). 
		%A cassette started with a lead plus stainless steel absorber (light blue and light grey) and ended at the Hexaboard of the second module. The cassette end was enclosed by a Mylar sheet, held by an aluminium frame (not represented here).
		%The zoomed-in area shows the first cassette. The lead plus stainless steel absorber of the second cassette is also shown including the air gap between two cassettes. The two drawings are proportionally scaled.}
	%\label{fig02:CEE-layout}
	%\end{figure*}

\subsection{2018 HGCAL prototype}
\label{sec:hgcalproto}

Full details of the detector construction and its readout electronics can be found in Refs.~\cite{beamtest-daq:2018,beamtest-calibration:2018}. 
In summary, the calorimeter prototype comprised three sections shown in \ref{fig:example_H2beamline}: the silicon electromagnetic calorimeter (CE-E), the silicon hadronic calorimeter (CE-H), and the tile hadronic calorimeter (AHCAL).
The electromagnetic calorimeter was constructed with \num{28} hexagonal detector modules, each of which was assembled as a glued stack, consisting of a sintered copper-tungsten baseplate, a silicon sensor, and a printed circuit board with the front-end readout electronics.
The hexagonal silicon sensor was sub-divided into 128 hexagonal pads, each with a surface of about \qty{1.1}{\cm\squared}.
To bias the sensor and shield it from electromagnetic interference, two metalized polyimide layers were used between the sensor and the base plate. 
%was the same as that used in our study of the EM performance of the calorimeter~\cite{beamtest-electrons:2018};
The \num{28} modules in the electromagnetic calorimeter were mounted in pairs on either side of copper cooling plates and these were interleaved with lead absorber plates.
%This is the same configuration that was used to study the electromagnetic performance of the detector reported in~Ref..
%Each 6" silicon sensor was part of modules were hexagonal in shape with an outer diameter of \qty{6}{\inch}. Each module had \num{128} silicon pads with an area of \qty{\sim1.1}{\cm\squared}.
%The CE-E section consisted of 28 sampling layers of hexagonal modules (6$^{\prime\prime}$ in diameter)
% with hexagonal Si pads ($\sim$ 1.1$cm{^2}$ per pad) interleaved with alternating
%copper and copper-tungsten absorbers or lead and stainless steel absorbers. 
%For the 2018 beam test, 28 hexagonal modules were assembled as a glued stack of a 
%%copper-tungsten baseplate, a polyimide foil, a silicon sensor and a readout PCB (referred to as Hexaboard). 
%
%The distance between the silicon layers varied between 8$mm$ -- 10$mm$. 
%This distance is also taken into account when calibrating the timing of the cells within the 
%whole detector. 
%Each hexagonal module was made as a glued stack consisting of a silicon sensor, which was sub-divided into 128 hexagonal pads ($\sim$ 1.1$cm{^2}$ per pad), a copper-tungsten base plate, a PCB with the front-end electronics mounted on it, and two polyimide layers used to bias the sensor and to shield from electronic noise. 
The first \num{26} layers of the electromagnetic section featured \qty{300}{\um} thick sensors, while the last two used \qty{200}{\um} thick sensors. 
The CE-E was approximately \qty{50}{\cm} long totaling about \num{28} radiation lengths, to ensure sufficient longitudinal containment of electromagnetic showers.
%A detailed description of the individual modules and the setup may be found in 

\subsubsection{Time measurement with the SKIROC2-CMS ASIC}
\label{subsec:sk2cms}

A 64-channel front-end readout ASIC was used to read out the modules. The ASIC, SKIROC2-CMS, was developed specifically for these tests by the OMEGA microelectronics group \cite{skiroc2-cms}.
The overall ASIC architecture and the data and control handling are described in detail in Refs.~\cite{beamtest-daq:2018,beamtest-calibration:2018}.
The ASIC measured both the amplitude and the time of arrival of the signals in each of the 64 channels. 
The signal amplitude was measured with two preamplifiers, with high and low gains, and a shaping time of \qty{40}{\ns}, whose outputs were digitized with separate Wilkinson ADCs. For the largest amplitude signals, a time-over-threshold (TOT) circuit was used. 
%The SKIROC2-CMS was designed and operated at the typical LHC collision frequency of \qty{40}{\MHz}. 

\begin{figure}[htb]
	\centering
	\includegraphics[width=0.52\linewidth]{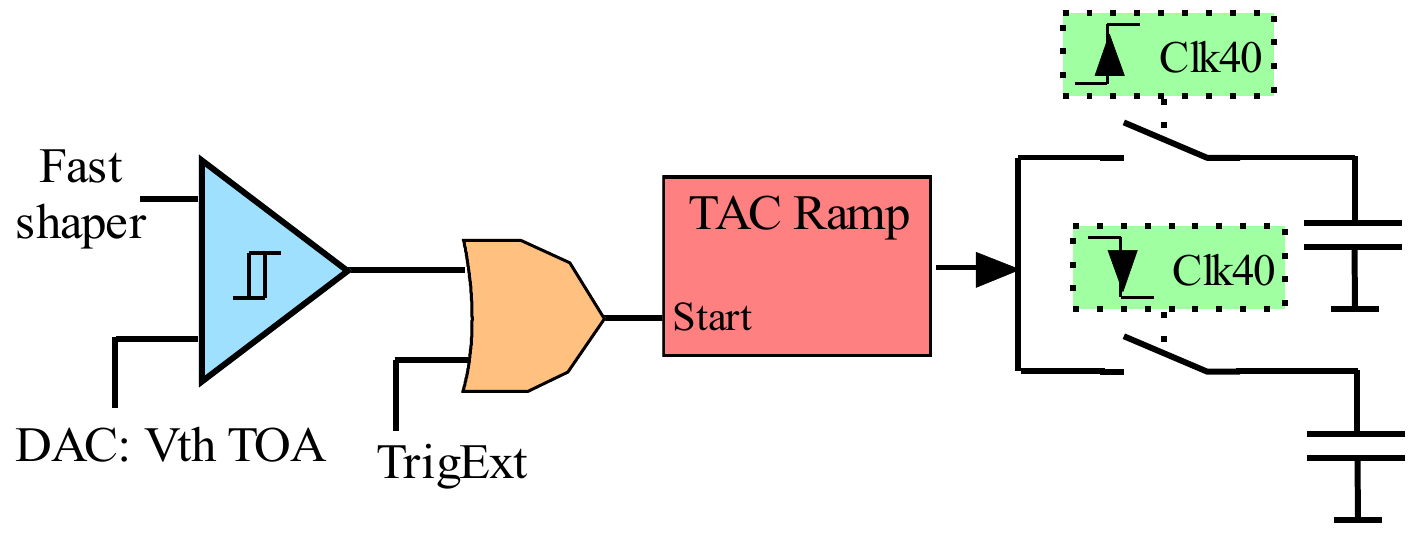}
    \hfill
	\includegraphics[width=0.43\linewidth]{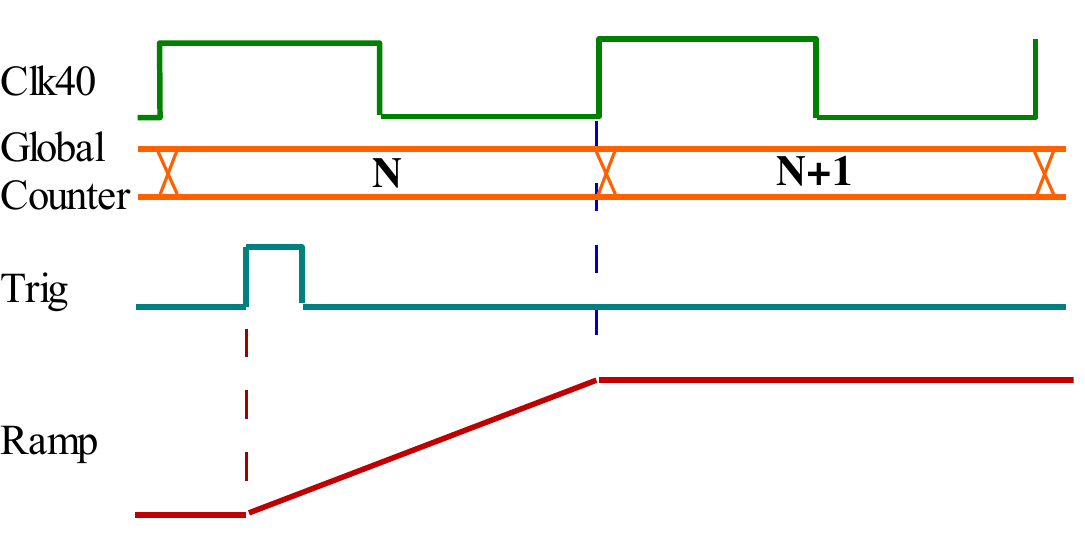}
	\caption{Block diagram of the SKIROC2-CMS time-of-arrival (TOA) analog circuit (left) and an illustration of the time-to-amplitude converter (TAC) ramp (right) that is stopped by a clock edge. An external trigger signal can be used to perform characterization and calibration of the TAC circuitry. Reproduced from Ref.~\cite{lobanov:2020}. 
}
	\label{fig:ch2_sk2cms_toa_block}
\end{figure}

The signal time was determined using a time-of-arrival (TOA) measurement relative to a common 40~MHz system clock.
The output of the preamplifier was fed into a fast shaper, with a shaping time of \qty{5}{\ns}, to remove high-frequency noise, and to shape the signal for input to a constant-threshold discriminator. 
The shaping time was chosen for optimal noise performance and to minimize time-walk effects, and the discriminator threshold was set to an energy corresponding to approximately \qty{15}{\MIP}, where \qty{1}{\MIP} corresponds to the average energy deposited by a minimum ionizing particle in the silicon sensor. 
%\footnote{The shapers for charge amplitude measurements had a \qty{40}{\ns} shaping time.}.
The output of the discriminator triggered two separate time-to-amplitude converter (TAC) circuits each with a voltage ramp, one of which was stopped on the subsequent rising edge of the clock, and the other on the falling edge, as illustrated in \ref{fig:ch2_sk2cms_toa_block}. 
%voltage ramps which were stopped upon arrival of the next-to-next clock edge. 
%There were two of such Time-to-Amplitude converter (TAC) ramps per channel.  
%
%The time relative to the clock was thus measured by so-called Time-To-Amplitude (TAC) converter ramps.
%The TACs used a voltage ramp that started with the discriminator trigger signal and stopped with the arrival of a clock edge.
% For each ASIC channel, there were two such TAC measurements, one stopping with the clock rising edge (ToA rise) and another stopping with the falling clock edge (ToA fall).
%
The first clock edge was skipped to avoid the strong non-linearity at the start of the voltage ramps.
%This led to an effective time range of \qtyrange{12.5}{37.5}{\ns}.
When the ramps were stopped, the voltage levels were stored in an analog memory and later digitized with a 12-bit Wilkinson ADC, similar to the one used in the gain measurements.
Due to the operation mode of the TACs, targeting the best possible time resolution of about \qty{50}{\ps}, the time measurement became non-linear for the last \qty{12}{\ns}. This was compensated by combining the rising and falling edge measurements, as described in \ref{sec:toa_recoCalib}.

%In addition, per each trigger frame a single timestamp was stored per ASIC indicating the clock cycle within which the first channel fired a ToA (cf.\ ``Global Counter'' in \ref{fig:ch2_sk2cms_toa_block}). This timestamp was not used since the information was too coarse to be useful.

\subsubsection{Clock path in the DAQ system}
%Details of the readout system can be found in Ref.~\cite{beamtest-daq:2018}.

\begin{figure*}[!ht]
	\begin{center}
		\includegraphics[width=1.\textwidth]{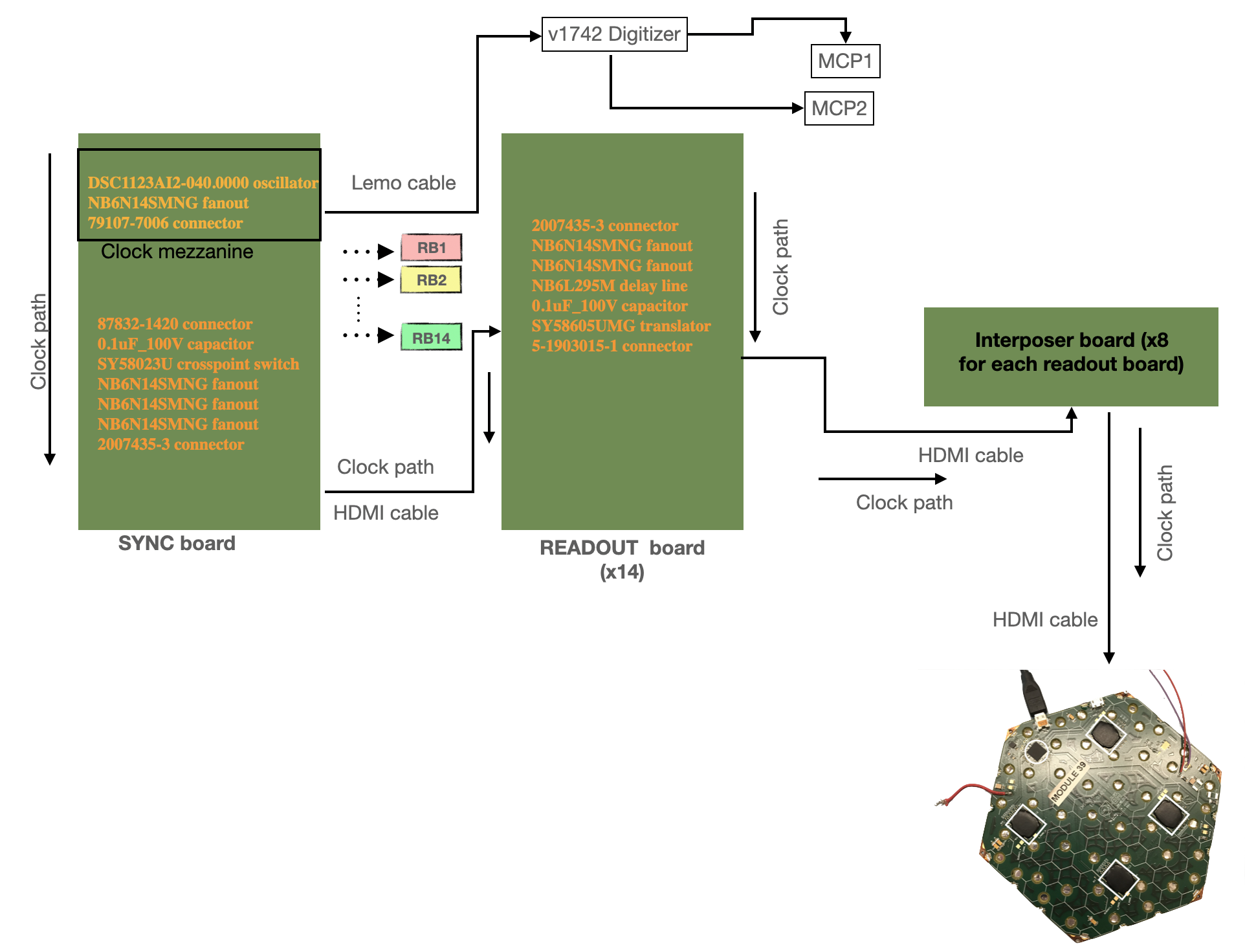}
	\end{center}
	\caption{Schematic view of the clock distribution tree in the experimental setup. All components on the clock path are explicitly mentioned. Laboratory measurements of the jitter from the synchronization board to the readout board outputs was measured to be less than \qty{10}{\ps} as expected.
	}
	\label{fig03:clockdist}
\end{figure*}

The \qty{40}{\MHz} system clock was generated with an oven-controlled quartz crystal oscillator (OCXO) %\todo{OCXO reference} 
on a custom-designed synchronization board (SB), which distributed the clock via HDMI cables to 14 custom-designed readout boards (RB). These, in turn, distributed the clock to seven detector modules via interposer boards, as shown in \ref{fig03:clockdist}.
The overall architecture is described in Ref.~\cite{beamtest-daq:2018}. 
A copy of the \qty{40}{\MHz} clock was also sent via an RG157 cable to a CAEN v1742 digitizer, which was also used to read out the analog signals from the two MCP detectors.

The jitter of the clock distribution system, from the clock-generating SB to the RBs, was measured in laboratory to be less than \qty{10}{\ps}.
However, the jitter of the clock between SB and the CAEN v1742 could not be determined in laboratory tests and was estimated to be \SI{50}{\pico\second} from \textit{in situ} measurements, as discussed below in \ref{sec:singleChannel_perf}.
%
%As the beam from the SPS beam was delivered in a continuous spill, the events recorded were asynchronous.
%The SB also distributed the trigger and control signals to all the RBs.

% On the SB, the \qty{40}{\MHz} clock was generated by a quartz crystal oscillator. There were several fanouts and connectors on the same synch board which lay in the path of clock signal. 
% This clock was then sent to the 14 RBs via HDMI cables. On each RB,
% there were various fanouts, connectors, and delay lines which again lay in the path of the clock signal. 

%Each RB then sent this clock to the 7 hexaboards via interposer boards~\cite{beamtest-daq:2018}. 

% In the presence of a coincidence signal from the two scintillators, a readout
% of the detector was triggered. The \qty{40}{\MHz} system clock, and trigger signals 
% were transmitted from the SB to the RBs via HDMI cables, and from there eventually to the module prototypes. 
%The SPS beam was delivered in a continuous spill and the beam particles were not synchronised with the clock. Hence, the trigger was asynchronous with respect to the beam signal. 

% Unfortunately, a jitter between SB and the MCP-reading CAEN v1742 could not be determined in laboratory tests.
% However, such a jitter can be determined in-situ from shower data, as discussed in \ref{sec:singleChannel_perf}.
% A consistent explanation of the various measured timing resolutions necessitates a MCP-SB jitter of about \SI{50}{\pico\second}.

\subsection{Datasets}
\label{sec:datasets}
%data
Data were collected with the HGCAL prototype with beams of positrons with energies between \qty{20}{\GeV} and \qty{300}{\GeV}, with approximately \num{50000} events collected at twelve different energies.

%Mention the data sample / size.
%E.g. only electron runs with energies 20-300 GeV. Total of N events available for calibration/analysis.

%MC
A \GEANTfour~\cite{AGOSTINELLI2003250} model of the experimental setup, including ancillary devices in the beam line, was used to simulate the detector response. 
This model is the same one used in Ref.~\cite{beamtest-electrons:2018}, where it is described in detail.
% The physics list for electromagnetic showers (FTFP\_BERT\_EMN) was used for the simulation of approximately one million particles in the same energy range as obtained from the beam test data.

% In this model, the signal time is %with the same effective approach used in Ref.~\cite{hgcal-tdr:2018}, where simulated hits from \GEANTfour are exploited to record 
% the generator-level time of the energy deposited in a detector cell: 
In order to obtain a realistic time resolution in the simulation, the time of the signal was convoluted with a Gaussian that had a width determined empirically for each channel.
Still, the model does not include the response of electronics components or the digitization, and thus any response non-linearities or similar effects are not simulated.

\section{Reconstruction of the timing information}
\label{sec:toa_recoCalib}
%\textcolor{red}{Primary: Thorben, Review: Artur}

\begin{figure}[htb]
	\centering
	\includegraphics[width=0.6\linewidth]{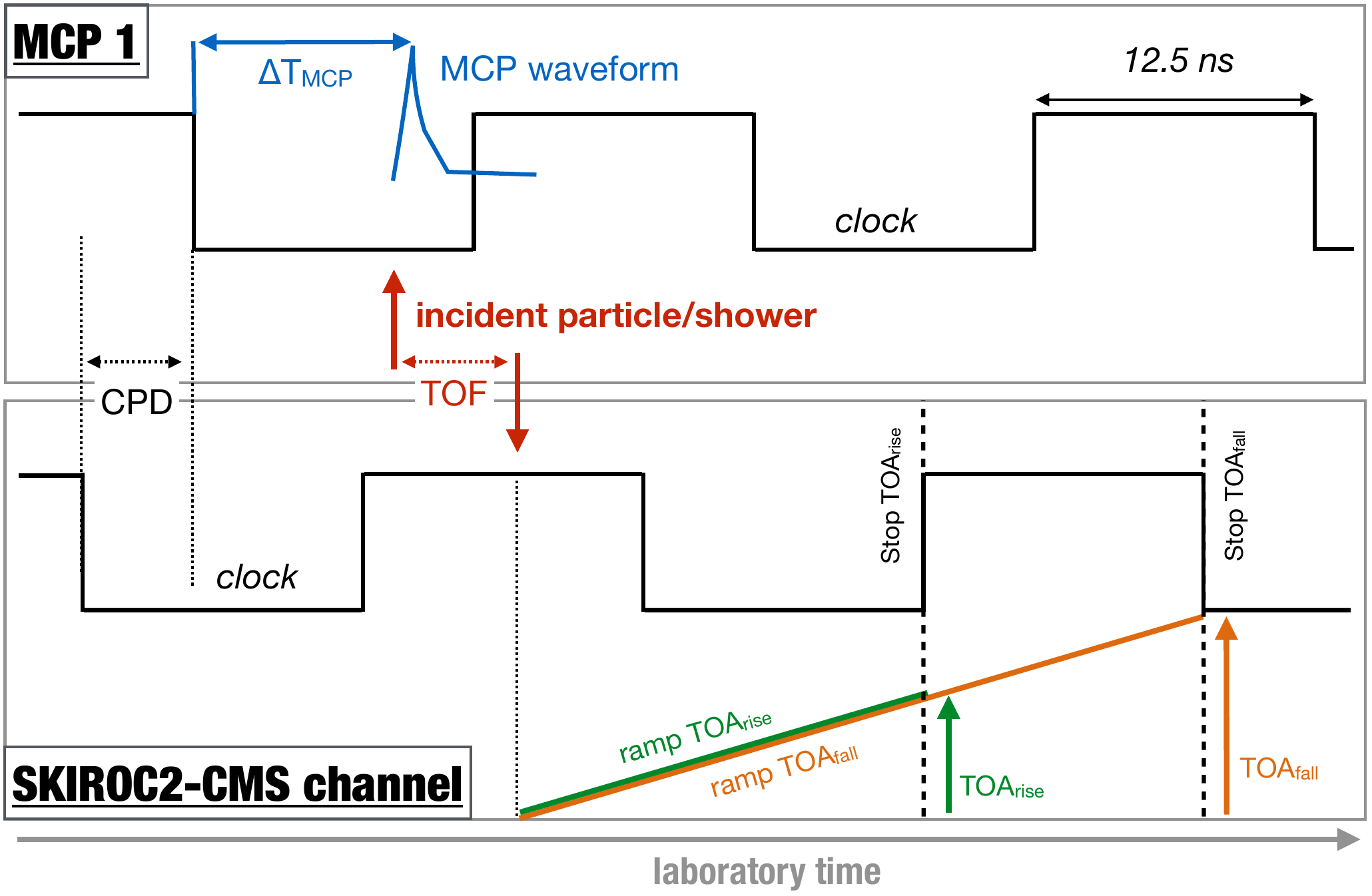}
	\caption{Timing measurements used in the calibration of the calorimeter hit timestamps. $CPD$ denotes an arbitrary but constant clock-phase-difference between a given readout channel and the reference clock. TOF refers to the time-of-flight between the MCP and silicon sensor.}
	\label{fig:timeinformation}
\end{figure}

The sequence of the time measurement in the ASIC is shown schematically in \ref{fig:timeinformation}.  
When the signal goes above a fixed threshold, the TAC ramps are started, and after skipping the first clock edge, the ramps are stopped: one on the rising edge of the clock, and the second one on the falling edge of the clock. This yields two time measurements, TOA-rise and TOA-fall, from which two separate signal times are estimated, $T_\text{rise}$\ and $T_\text{fall}$. These times are referenced to the system clock after corrections for the non-linearity of the TAC ramp, the time-walk that depends on the hit energy, and the deposited energy in the full module. 
% These are the time of the energy deposition in a readout channel
% without any correction for time-of-flight between the layers or for variations due to angle of incidence. 
%irrespective of its location\footnote{Defining as the time of incidence onto the calorimeter, \ie irrespective of a readout channel's location, has a negligible impact on performance results.}.

\subsection{Signal time reconstruction}
To estimate the signal time, $T$, the procedure for both TOA measurements was as follows: 

\begin{enumerate}
    \item The measured TOA values were normalized to the unit interval to take into account their pedestal values,
	\item the TOAs were corrected for the non-linearity of the TAC ramp ($f_\text{TOA}$),
	\item an amplitude-dependent time walk correction ($f_\text{TW}$) was applied, and
	%The dependence of the TOA measurement on the reconstructed hit energy ($E_\text{hit}$), used as proxy for the signal amplitude, that is induced by the fixed-threshold discrimination. This effect is known as time walk and is denoted as $f_\text{TW}$ in this work.
    \item a correction for small signals ($f_\text{R}$) that depends on the total energy deposited in the module containing the channel ($E_\text{module}$) was applied.
\end{enumerate}
It is worth noting that the first two steps operate on TOA values and their distribution, whereas the last two steps operate on individual signal times.
Accordingly, the signal time, $T$, for each channel has three contributions, as shown in \ref{eq:toa_to_time}:
\begin{equation}
	T_\text{rise/fall}~=~f_\text{TOA}(TOA_\text{rise/fall})~+~f_\text{TW}(E_\text{hit})~+~f_\text{R}(E_\text{module}, E_\text{hit}) .
	\label{eq:toa_to_time}
\end{equation}
The signal time $T$ in any given channel is estimated separately from both TOA-rise and TOA-fall. 
%The calibration procedure described in \ref{subsec:calibration} is the same for both TOA-rise and TOA-fall measurements and in the following section no distinction is made between them.

\subsection{Derivation of the corrected time}
\label{subsec:calibration}
As the beam particles were asynchronous with respect to the \qty{40}{\MHz} system clock, the signal arrival times were uniformly distributed within the \qty{25}{\ns} clock period. 
This resulted in the full range of possible TOA values being available for the determination of the non-linearity corrections. 

%This resulted in the full range of possible TOA values being used, and thus required  the non-linearity correction and the combination of both TOA measurements, rise and fall to achieve the best possible time response and resolution.

%As the time difference between particle time and the TOA is independent of its location in a clock period and only depends on the time-of-flight (TOF) and signal propagation differences, which are constant, the uniformity of the signal arrival time was used to derive the non-linearity corrections to the TOA values.  
  
% The time calibration of a readout channel is obtained by fitting the from \ref{eq:timestampdef}.
% For each channel and for the rise and fall variants separately, the expression in \ref{eq:toa_to_time} is fit to the sum of the two terms in \ref{eq:timestampdef}: the time of the signal in the MCP ($\Delta T_\text{MCP}$) and the time-of-flight (TOF) from the MCP to the readout channel's location.
%The TOF is estimated from the beam particle trajectory ($\Delta X$), reconstructed with wire chambers placed upstream of the MCP, and assuming that the shower particles proceed through the detector prototype at the speed of light ($c$).
% \begin{equation}
% 	T~~\leftrightarrow~~\Delta T_\text{MCP}~+~TOF ~=~ \Delta T_\text{MCP}~+~\frac{\Delta X}{c}
% 	\label{eq:timestampdef}
% \end{equation}
The signal from MCP1 was required to have an amplitude greater than \qty{500}{ADC} counts, where the time resolution was better than \qty{40}{\ps}, as shown in \ref{fig:mcpreso}.
%The times computed from MCP1 were aligned with the TOA readings in a given channel by correcting for the \qty{25}{\ns} periodicity of the digital clock, cf.~Figure~9.19a in Ref.~\cite{Quast:2020}.
Also, only readout channels with more than \num{1000}~hits with $E_\text{dep}\geq\qty{250}{\MIP}$ and with \num{30000} or more hits in total were considered.
The first requirement selects a large sample of measurements where the energy is estimated from the TOT measurement, i.e.\ not from the ADC measurement.
As the beam was focused on the center of the calorimeter, only \num{116} readout channels, or 3\% of all channels, met these requirements.

\subsubsection{Correction of the non-linearity of the TOA}
\label{subsubsec:toa_lin}
 
First, variations in pedestals ($TOA^\text{min}$) were corrected, and the values were scaled by the full range ($\Delta TOA$), yielding $TOA_\text{norm}$ normalized to unity, thus corresponding to the relative location of the TOA value in the clock period:
\begin{equation}
	TOA_\text{norm} ~\coloneqq~ \frac{TOA-TOA^\text{min}}{\Delta TOA}.
	\label{eq:toa_norm}
\end{equation}

After this normalization, the non-linear response of $TOA_\text{norm}$ was corrected using the MCP measurement, which was considered linear.
For each channel the response was modelled using
\begin{equation}
	{f}_{TOA}(TOA_\text{norm} | \vec{\Theta}) ~=~\Theta_1 \cdot x + \Theta_2 + \frac{\Theta_3}{x-\Theta_4},
	\label{eq:toa_tw_param}
\end{equation}
where $\vec{\Theta}$ is a set of parameters describing the response of the TOA measurement.
%derived from the fit to the distribution of signal times in a clock period.
%with the $\Theta_2$ corresponding to an offset constant, denoted as $CPD$ in \ref{fig:timeinformation}.

%After this normalization, \ref{eq:toa_tw_param} was fit to the average of the MCP signal times as a function of $TOA_\text{norm}$. 
The accuracy of this response linearization is improved by using separate $\vec{\Theta}$ parameter sets in the linear ($TOA_\text{norm} < 0.65$) and non-linear regions ($TOA_\text{norm} \geq 0.65$).

%\begin{equation}
%	f_\text{TOA}(TOA_\text{norm}) ~\coloneqq~
%	\begin{cases}
%		\hat{f}(TOA_\text{norm}|\vec{\Theta}_{1}^{TOA})&\text{for }TOA_\text{norm} < 0.65, \\
%		\hat{f}(TOA_\text{norm}|\vec{\Theta}_{2}^{TOA})&\text{for }TOA_\text{norm} \geq 0.65.
%	\end{cases}	
%	\label{eq:toa_calib}
%\end{equation}

\begin{figure}
% 	\captionsetup[subfigure]{aboveskip=-1pt,belowskip=-1pt}
	\centering
	\begin{subfigure}{0.325\textwidth}
% 		\centering
		\includegraphics[width=\textwidth]{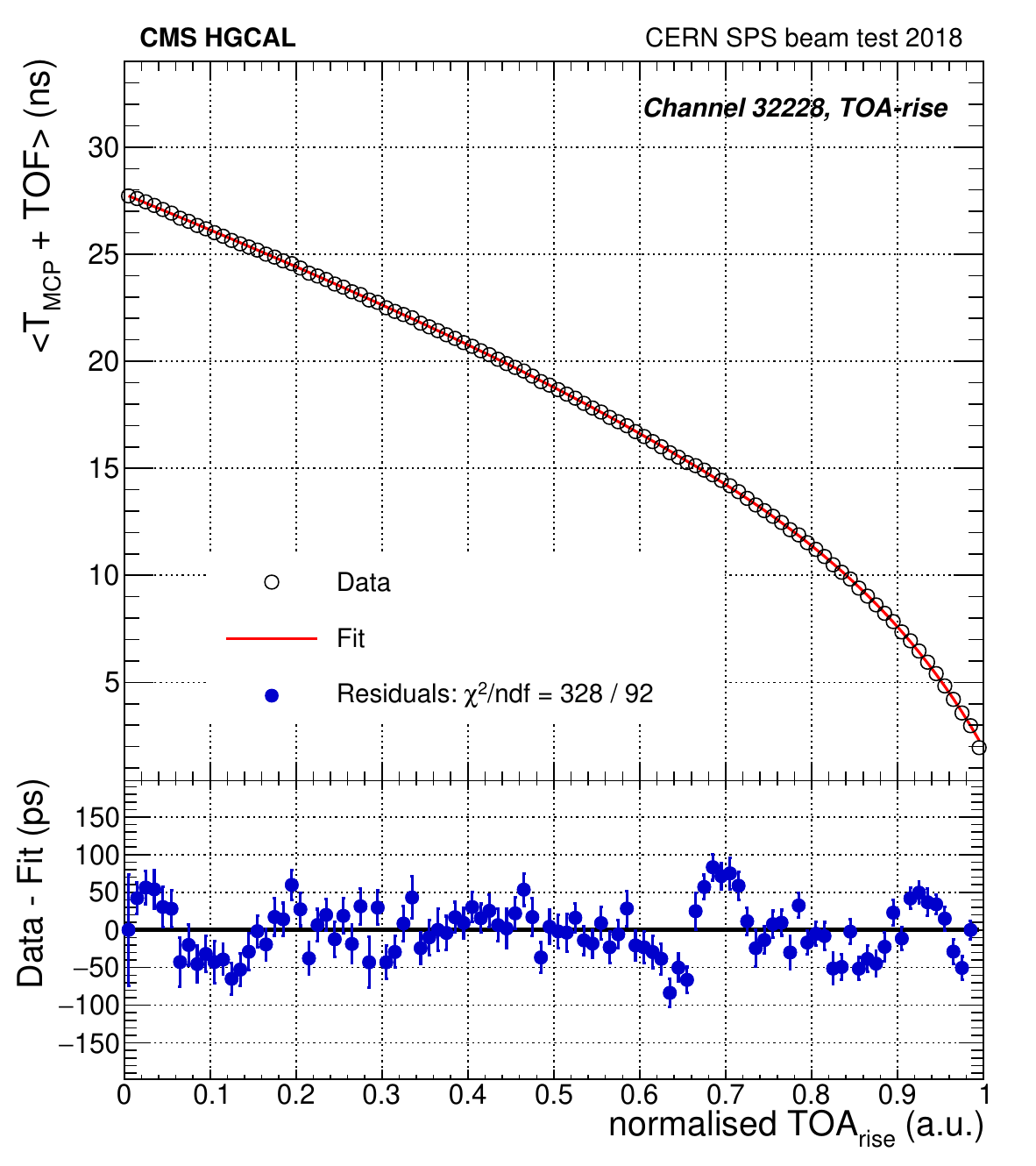}
		\subcaption{TOA linearization, \ref{subsubsec:toa_lin}.
		\label{plot:TOA_calib_rise}
		}
	\end{subfigure}
% 	\hfill
	\begin{subfigure}{0.325\textwidth}
% 		\centering
		\includegraphics[width=\textwidth]{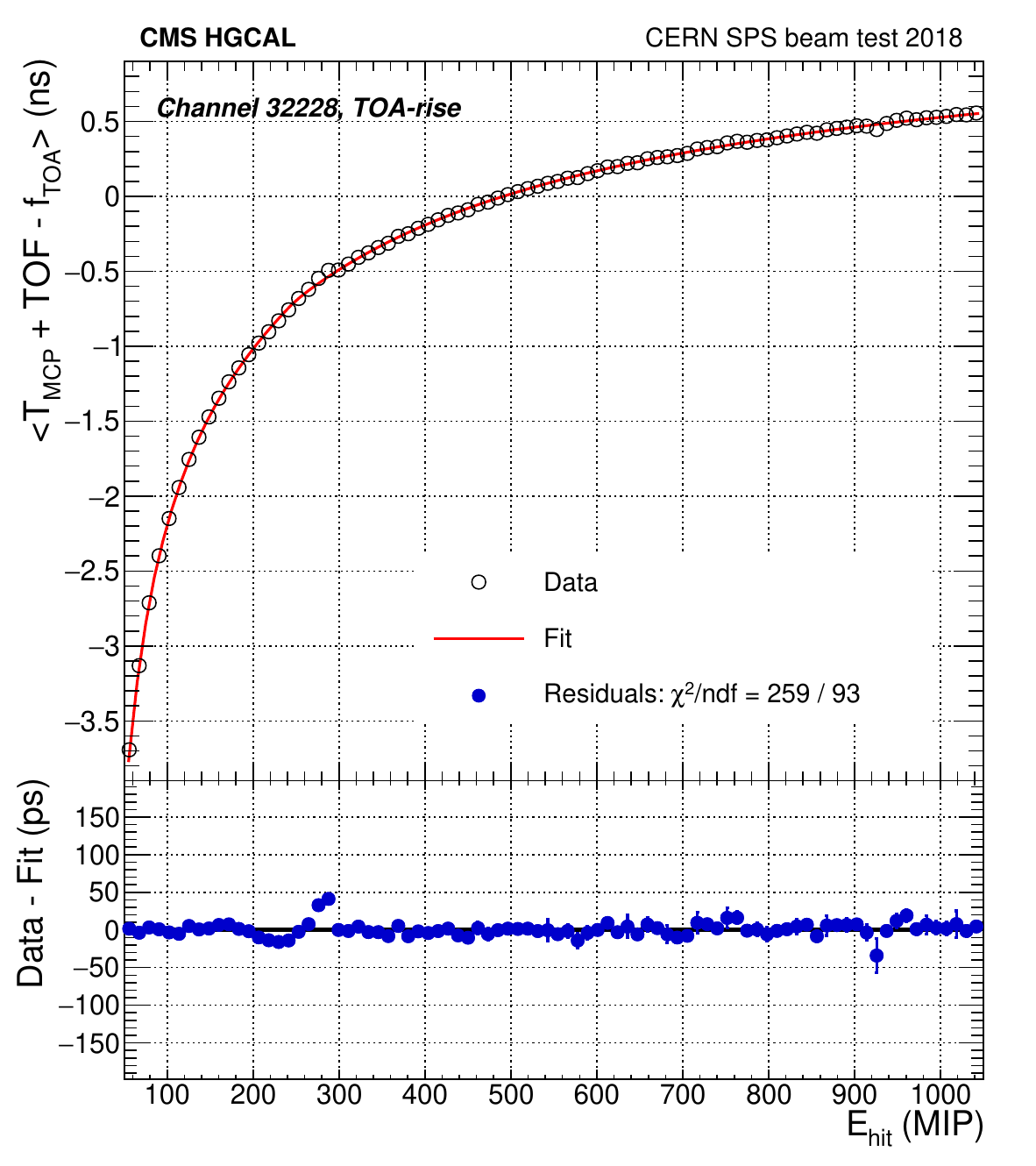}
		\subcaption{Time-walk correction, \ref{subsubsec:tw}.
		\label{plot:TW_calib_rise}
		}
	\end{subfigure}
% 	\hfill
	\begin{subfigure}{0.325\textwidth}
% 		\centering
		\includegraphics[width=\textwidth]{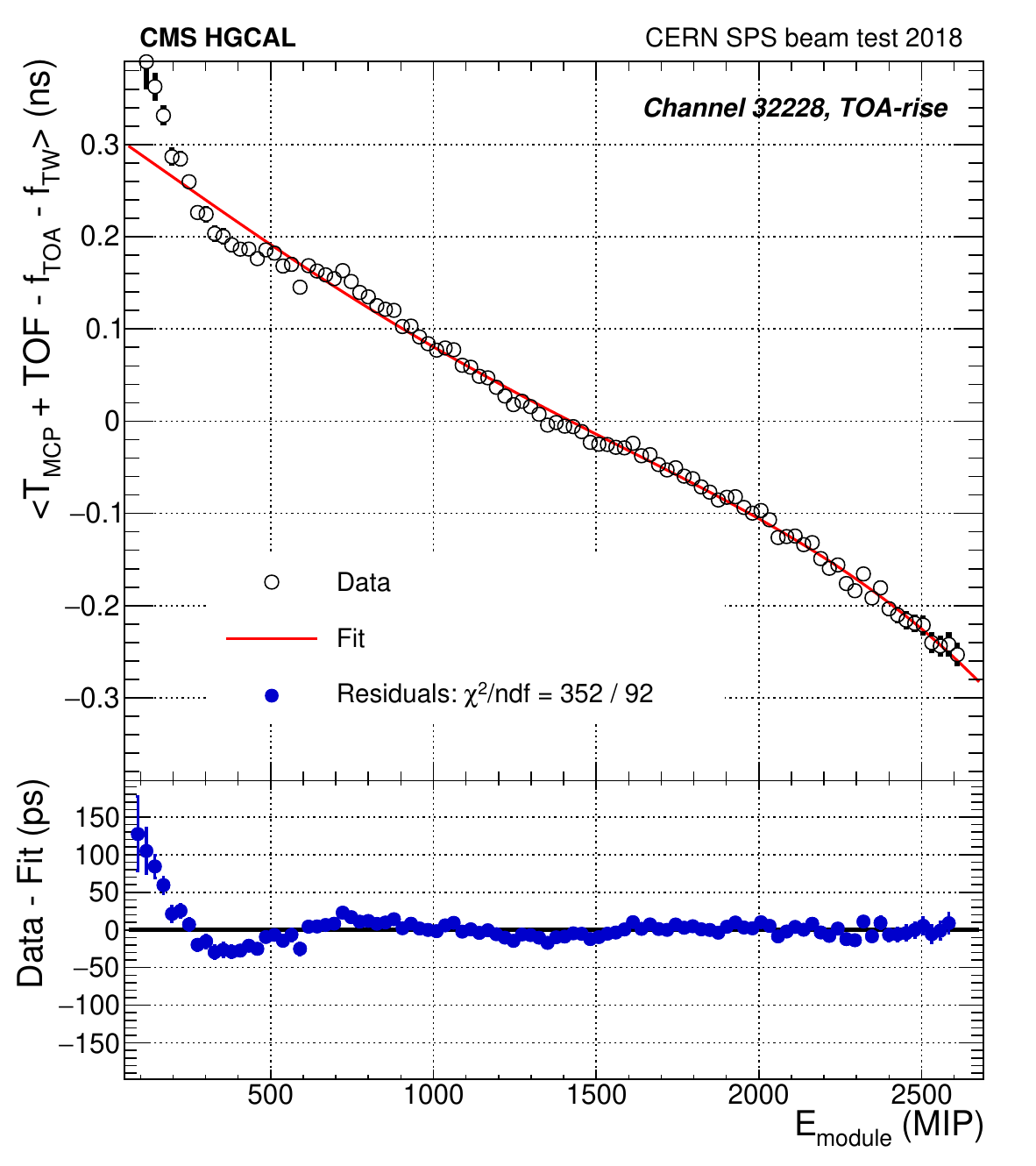}
		\subcaption{Residual correction, \ref{subsubsec:residual}.
		\label{plot:ESUMTW_calib_rise}
		}
	\end{subfigure}	
	\caption{
		The three TOA calibration steps (TOA-rise in this case) of a representative channel, centrally located inside the HGCAL prototype:
		(a) Linearization of the normalized TOA,
		(b) assessment of the signal-induced time-walk correction, and
		(c) assessment of the residual correction, a smaller time-walk that depends on the total energy in the module of the given channel.
		The magnitude of the time-walk corrections is about one and two orders of magnitude smaller than the calibrated time range, respectively.
		The additive TOF term represents the constant time offset between the MCP and the readout channel.
	\label{plot:calibration_steps}
	}
\end{figure}

The result of the linearization step is shown in \ref{plot:TOA_calib_rise} for the normalized TOA-rise of a representative channel, where the full \qty{25}{\ns} range is presented.
This method is found to be in agreement with the previous response correction presented in Ref.~\cite{lobanov:2020}, that did not rely on the MCP reference but on the asynchronous nature of the beam particles.
% , whereas this method allows to calibrate more channels due to the less strict selection requirements.
%It is found that the parameterizations thus derived are consistent with the results from the first calibration results based on the asynchronous nature of the beam particles~\cite{lobanov:2020}.

\subsubsection{Time-walk correction}
\label{subsubsec:tw}
The $f_\text{TOA}$ linearization corrections were followed by an amplitude-dependent correction for the time-walk effect. This correction is derived from a fit to the TOA time difference to the MCP as a function of the reconstructed signal amplitude, $E_\text{hit}$ using the same functional form as in \ref{eq:toa_tw_param}.
% that is used as proxy for the signal magnitude.
% Therefore, the model is fit to the average of the reference timestamps corrected by $f_\text{TOA}$, as a function of $E_\text{hit}$.
It was also found that the fit was improved by separately fitting two regions of the signal amplitude depending on whether the ADC ($E_\text{hit} < E_\text{TOT}$) or the TOT is used ($E_\text{hit} \geq E_\text{TOT}$) to estimate the hit energy:
\begin{equation}
	f_\text{TW}(E_\text{hit}) ~\coloneqq~
	\begin{cases}
		{f}_{TOA}(E_\text{hit}|\vec{\Theta}_{1}^{TW})&\text{for }E_\text{hit} < E_\text{TOT}. \\
		{f}_{TOA}(E_\text{hit}|\vec{\Theta}_{2}^{TW})&\text{for }E_\text{hit} \geq E_\text{TOT}. 
	\end{cases}	
	\label{eq:toa_twcalib}
\end{equation}
The value of $E_\text{TOT}$ was determined channel-by-channel and is typically a few hundred MIPs \cite{beamtest-calibration:2018}.
As can be seen in \ref{plot:TW_calib_rise}, the time-walk is found to reach several nanoseconds for hit energies below a few \qty{100}{\MIP}.

\subsubsection{Residual time-walk correction depending on module energy}
\label{subsubsec:residual}
After the linearization and time-walk corrections, timing corrections of the order of a few \qty{100}{\ps} were needed for small energies. This effect depended on the total energy deposited in the module that channel belonged to, $E_\text{module}$. It is likely that this effect is due to the common-mode estimation procedure~\cite{beamtest-calibration:2018} that is applied only to small signals measured with the ADC signals and not to large ones measured with the TOT. 
We found that this residual correction can be well modeled by
\begin{equation}
	f_\text{R}(E_\text{module}, E_\text{hit})~\coloneqq~
	\begin{cases}
		\mathcal{P}_4(E_\text{module})&\text{for }E_\text{hit} < E_\text{TOT}, \\
		0&\text{for }E_\text{hit} \geq E_\text{TOT},
	\end{cases}	
	\label{eq:toa_twmodule}
\end{equation}
where $\mathcal{P}_4$ represents a fourth-degree polynomial whose parameters are determined from a fit to the average of the reference timestamps corrected by $f_\text{TOA}+f_\text{TW}$, as a function of $E_\text{module}$.
\ref{plot:ESUMTW_calib_rise} shows the residual correction determined for a representative channel.
%\newline
%\textcolor{red}{Artur: please review and improve the following explanation.}

% This residual correction may in principle be due to the procedure used to estimate and subtract common-mode fluctuations~\cite{beamtest-calibration:2018} that coherently affect a large number of channels in the same module. This procedure only applies to ADC data and, therefore, to small signals.  
% For larger signals, namely when $E_\text{hit}>E_\text{TOT}$, it is observed that the residual correction is negligible, as directly reflected in  \ref{eq:toa_twmodule}. For these hits, measured with the TOT, the common-mode correction does not apply, lending support to the hypothesis that the common-mode subtraction procedure may be at the root of the need for the residual correction.
% While identifying the cause for this effect is beyond the scope of this work, it is suspected that the need for this residual correction is specific to the reconstruction of signals in the SKIROC2-CMS ASIC and does not necessarily generalise to other ASICs.

\subsubsection{Combination of TOA-rise and TOA-fall}

If both TOA-rise and TOA-fall values are within their linear region, the time is computed from the average of the two timing estimates. 
Otherwise, the variant in, or closest to, its linear region is retained.

After all corrections were applied the estimated precision of the time measurement among the 116 channels for which the time calibration procedure was possible is about \qty{50}{\ps}.

The fitted calibration parameter values for the example channel illustrated in \ref{plot:calibration_steps} are provided in \ref{appendix:calib_constants}.
The parameter values vary by about 10\% over the calibrated channels.
The analysis of the timing performance of the detector discussed below is restricted to those channels, that are located centrally in the prototype.

\begin{figure}
% 	\captionsetup[subfigure]{aboveskip=-1pt,belowskip=-1pt}
% 	\centering
	\begin{subfigure}{0.249\textwidth}
% 		\centering
		\includegraphics[trim=58 70 67 41,clip,width=\textwidth]{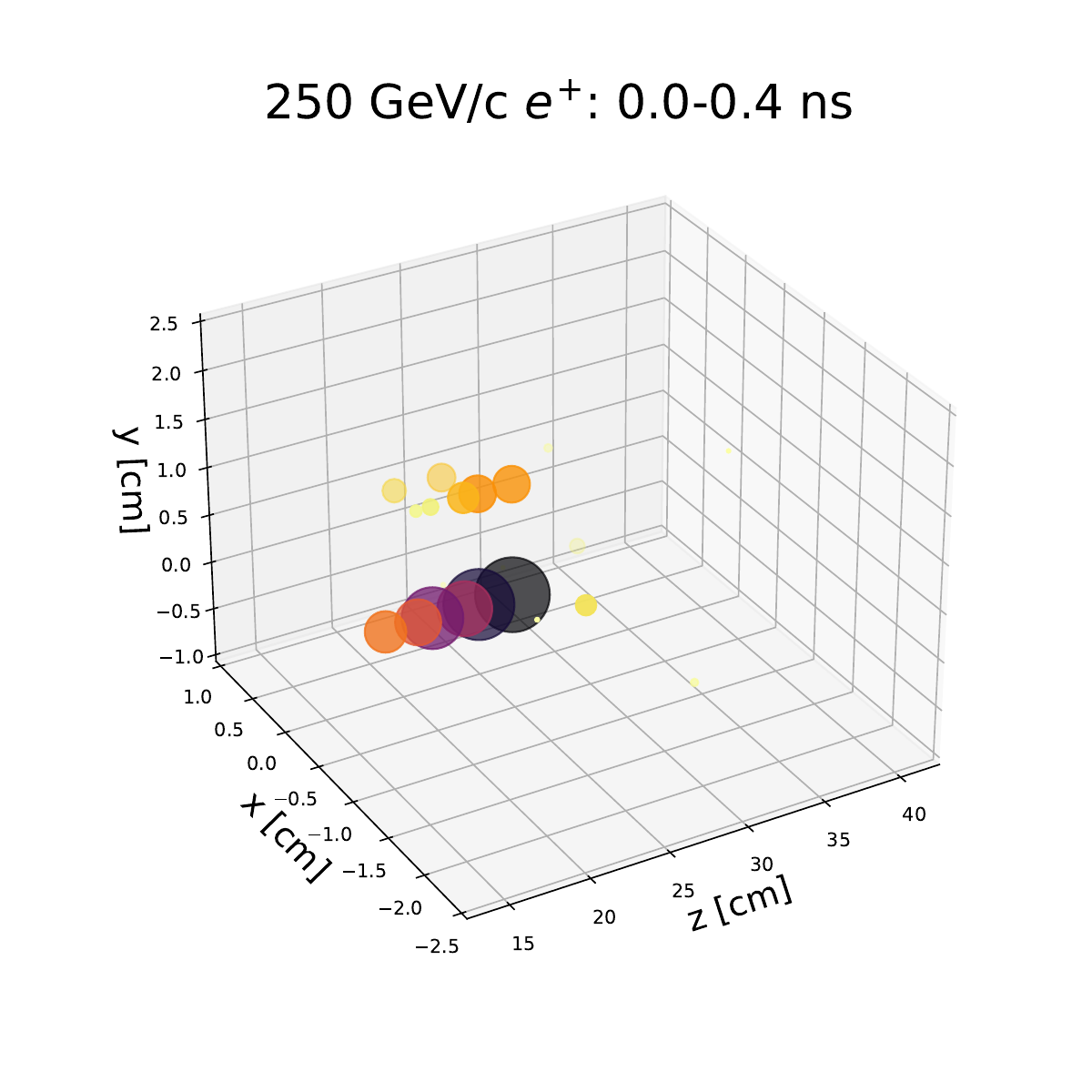}
		\subcaption{$\qty{0}{\ns}\leq T\leq\qty{0.4}{\ns}$.
		\label{plot:event_video_ts1}
		}
	\end{subfigure}
% 	\hfill
    \hspace{-2mm}
	\begin{subfigure}{0.249\textwidth}
% 		\centering
		\includegraphics[trim=58 70 67 41,clip,width=\textwidth]{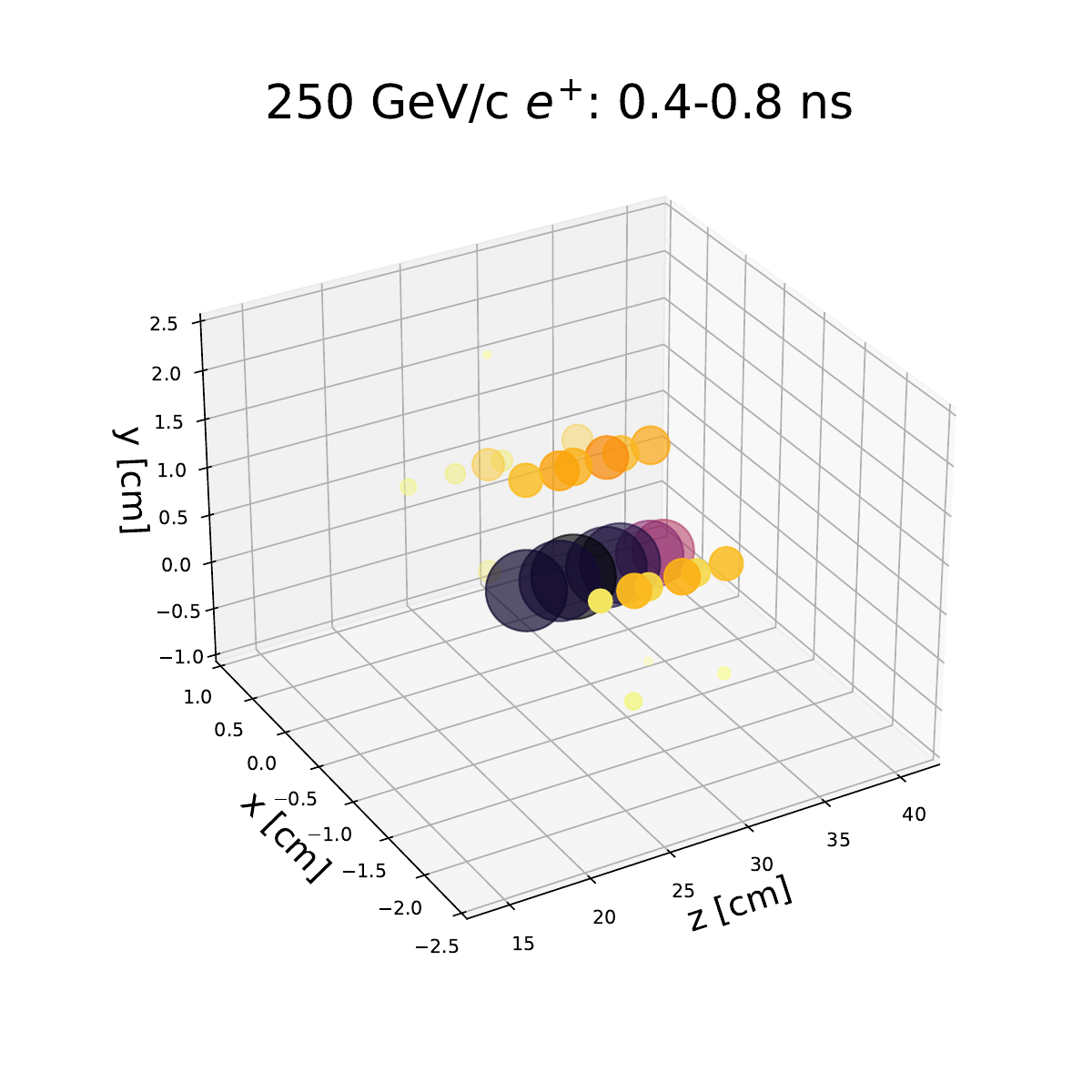}
		\subcaption{$\qty{0.4}{\ns}<T\leq \qty{0.8}{\ns}$.
		\label{plot:event_video_ts2}
		}
	\end{subfigure}
% 	\hfill
    \hspace{-2mm}
	\begin{subfigure}{0.249\textwidth}
% 		\centering
		\includegraphics[trim=58 70 67 41,clip,width=\textwidth]{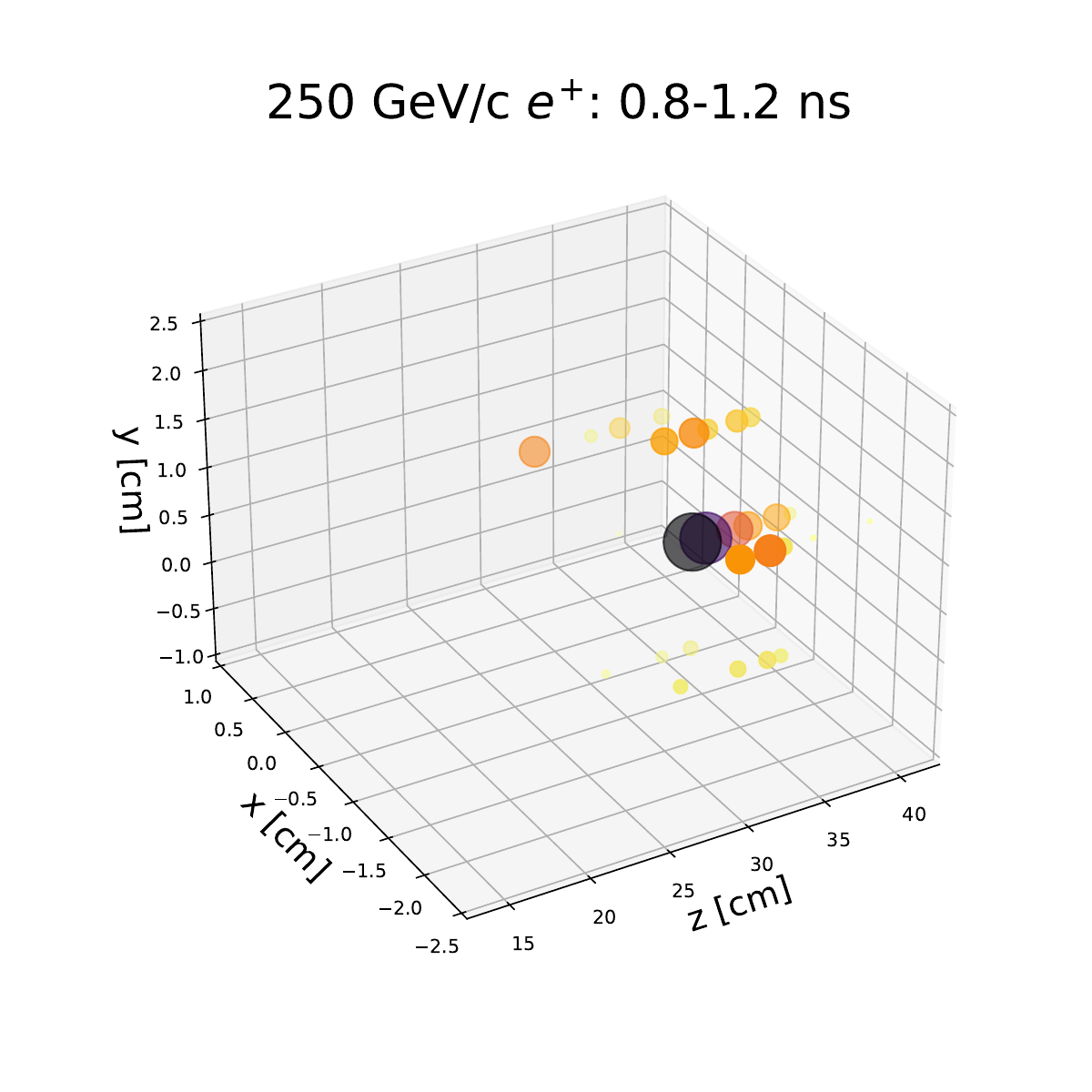}
		\subcaption{$\qty{0.8}{\ns}<T\leq \qty{1.2}{\ns}$.
		\label{plot:event_video_ts3}
		}
	\end{subfigure}	
% 	\hfill	
    \hspace{-2mm}
	\begin{subfigure}{0.249\textwidth}
% 		\centering
		\includegraphics[trim=58 70 67 41,clip,width=\textwidth]{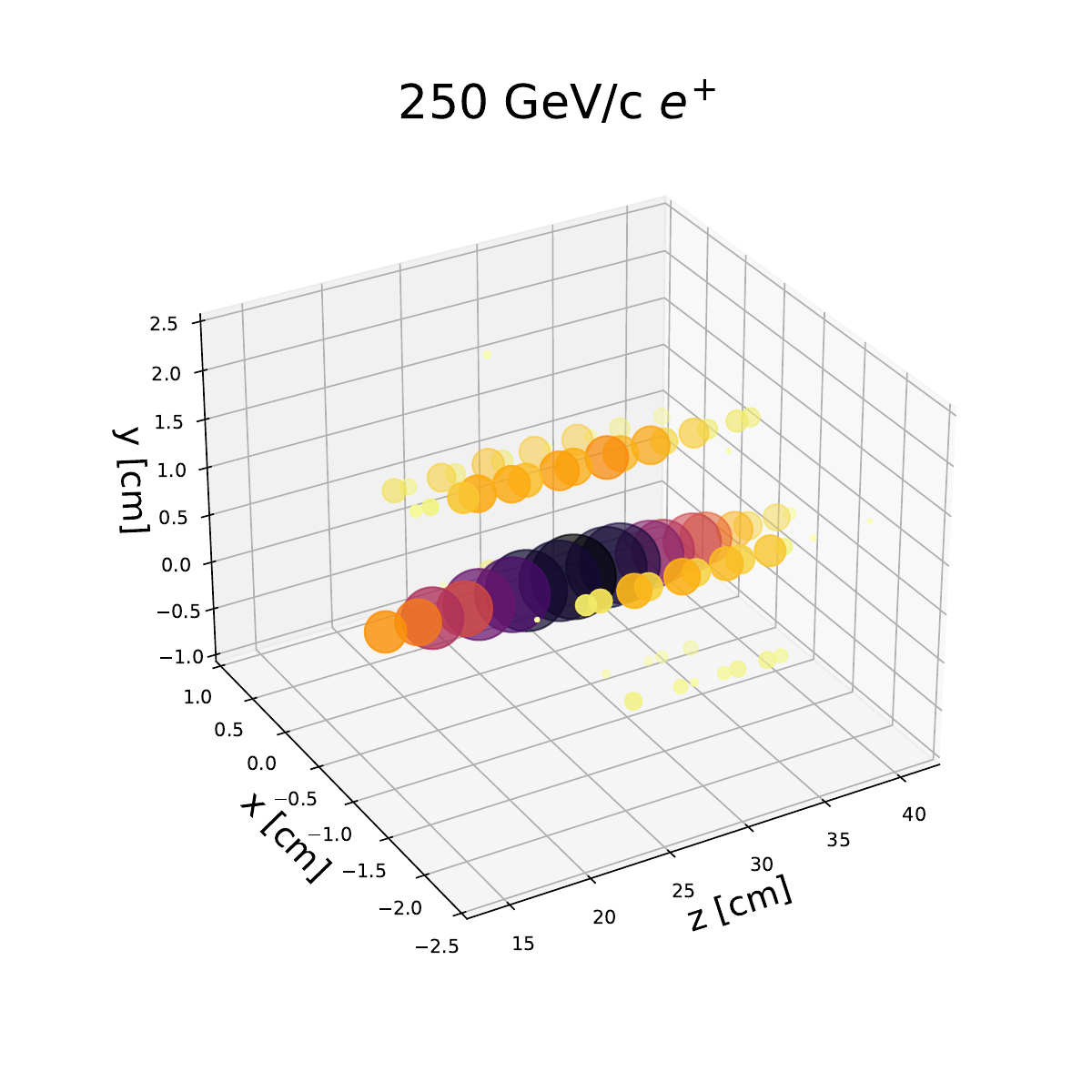}
		\subcaption{Full shower.
		\label{plot:event_video_incl}
		}
	\end{subfigure}		
	\caption{
		Time evolution (a--c) of a shower (d) induced by a \qty{250}{\GeV} positron in the HGCAL prototype.
		Data are shown for \num{81} hits with reconstructed timestamps in this event.
		Hit energies are proportional to larger and darker markers, and are provided solely for illustration purposes.
	}
\end{figure}

To illustrate the results of the timing reconstruction and calibration procedure, \ref{plot:event_video_ts1,plot:event_video_ts2,plot:event_video_ts3} show the time evolution of the hits from a \qty{250}{\GeV} positron showering in the calorimeter. 
A subset of the calibrated channels is visible along the core of the shower, and a clear correlation is observed between the reconstructed time and the spatial development of the shower, as expected.
For this event, the reconstructed hit energies vary by a factor of about 20.

\section{Timing performance of single channels and full showers in the HGCAL prototype}

In this section we describe the time performance from the single channel level to full showers in the prototype.
First, the time resolution is determined for individual channels using the MCPs as an external reference.
Then, this resolution model is injected into the simulation and compared to the shower timing as measured in the particle data. Further studies of the HGCAL-only full shower time resolution conclude this section.

\subsection{Single channel performance}
\label{sec:singleChannel_perf}
%Content:  - calibrated ToA vs cell energy with MCP as reference,
%                           - calibrated ToA vs cell energy with a Si channels as reference. [This in 
%                             comparison with the previous point and the MCP-MCP performance in 
%                             chapter 3. allows to introduce the global jitter concept] 
%                           - summary statistics (jitter and constant term) of per-channel timing 
%                                       resolution vs. energy)
%                           - calibrated ToA vs cell energy with MCP as reference, considering the 
%                              difference to the average time of the shower. [Analysis of the data 
%                              accounting for the presence of the extra jitter,  and its quantification 
%                              through comparison to the first point]

%\textcolor{red}{Responsible: Axel, contributors / reviewers: Arabella, Axel, Thorben}

The per-cell timing performance is a fundamental ingredient to build a realistic simulation with which one can gauge the shower timing measurements obtained in data.

\begin{figure}[hbtp!]
	% 	\captionsetup[subfigure]{aboveskip=-1pt,belowskip=-1pt}
		\centering
		\hfill
		\begin{subfigure}[b]{0.4\textwidth}
			\centering
			\includegraphics[width=\textwidth]{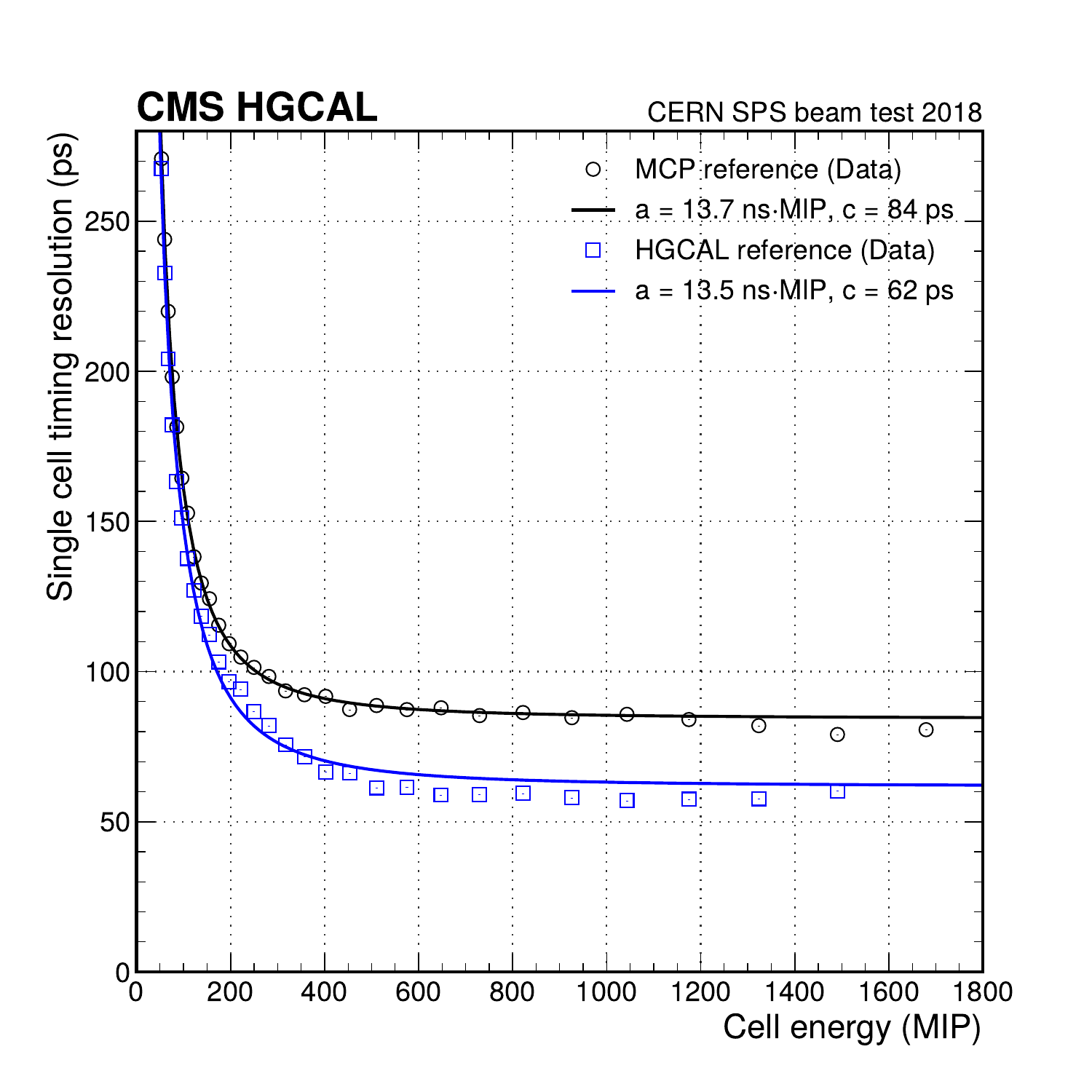}
			\subcaption{Single-channel timing performance taking the MCP (black circles) or the HGCAL shower measurement (blue squares) as reference.
			\label{fig:tReso_vsCell_MCP_HGC_ref}
			}
		\end{subfigure}
		\hfill
		\begin{subfigure}[b]{0.5\textwidth}
			\centering
			\includegraphics[width=\textwidth]{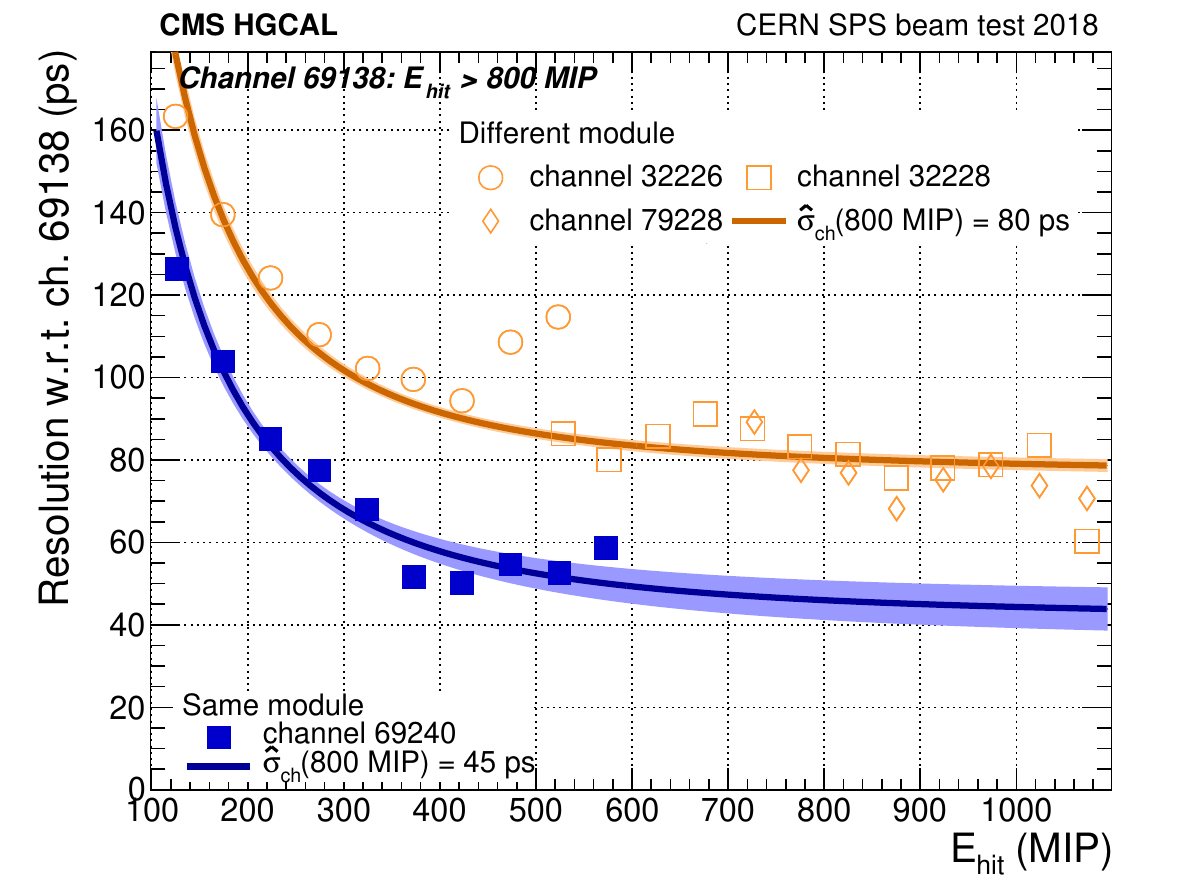}
			\subcaption{Single-channel timing performance taking as the reference a different silicon channel from a different module (orange, open) or from a different ASIC on the same module (blue, filled).
			\label{fig:tReso_vsCell_Siref}
			}
		\end{subfigure}
	%	\begin{subfigure}[b]{0.3\textwidth}
	%		\centering
	%		\includegraphics[width=0.999\textwidth]{plots/Chapter5/timing_resolution_vs_cell_energy/sigmas_vs_cell_energy_electron.pdf}
	%		\subcaption{MCP reference.
	%		}
	%		\label{fig:tReso_vsCell_MCPref}
	%	\end{subfigure}
	%	\hfill
	%	\begin{subfigure}[b]{0.3\textwidth}
	%		\centering
	%		\includegraphics[width=0.999\textwidth]{plots/Chapter5/timing_resolution_vs_cell_energy/sigmas_minus_shower_average_vs_cell_energy_electron.pdf}
	%		\subcaption{Shower time as reference.
	%		}
	%        \label{fig:tReso_vsCell_HGCref}
	%\end{subfigure}
		\hfill
		\caption{Single-cell time resolution as a function of the cell energy in data, computed against different timing references:
		In (a) the MCP measurement or the average time of the shower measured with the HGCAL prototype are used.
		The difference in performance between the MCP and HGCAL references is evidence of a jitter between the two systems.
		In (b) a reference silicon channel is used, that is chosen from a different module or from a different ASIC of the same module.
		The difference in performance between the same-module and different-module references is evidence of a correlation between the timing measurements of different channels.
		\label{fig:tReso_vsCell_MCP_Si}
		}
	\end{figure}

The average per-cell timing performance of the calibrated TOA values is quantified as a function
of the corresponding energy deposit.
For this purpose, the TOA values are compared to a reference time measurement in bins of energy,
and fitted using a Gaussian resolution function.
The measured average time is found to be rather flat as a function of the energy, with deviations
up to about \qty{20}{\ps} in the very low energy region (below \qty{300}{\MIP}), that are 
consistent with the outcome of the calibration procedure described in \ref{sec:toa_recoCalib}.
The measured resolution as a function of the deposited energy is shown in \ref{fig:tReso_vsCell_MCP_Si}
for two different choices of reference time measurement: in \ref{fig:tReso_vsCell_MCP_HGC_ref} (black) the reference time is provided by the MCP system, while in \ref{fig:tReso_vsCell_Siref} it is provided by one silicon cell within the HGCAL.
 In \ref{fig:tReso_vsCell_Siref}, results are shown for pairs of cells from different modules (orange) and in the same module and different ASICs (blue).
The measurements in \ref{fig:tReso_vsCell_MCP_Si} are fitted with $\sigma^2(E) = (a/E)^2 + c^2$, where $E$ represents the energy,
$a$ is a constant that represents the improvement in resolution with energy, and $c$ is a constant term.

%\begin{figure}[hbtp!]
%        \captionsetup[subfigure]{aboveskip=-1pt,belowskip=-1pt}
%        \centering
%        \hfill
%        \begin{subfigure}[b]{0.43\textwidth}
%                \centering
%                \includegraphics[width=0.999\textwidth]{plots/Chapter5/timing_resolution_vs_cell_energy/sigmas_vs_cell_energy_electron.pdf}
%                \subcaption{MCP reference.
%                }
%                \label{fig:tReso_vsCell_MCPref}
%        \end{subfigure}
%        \hfill
%        \begin{subfigure}[b]{0.53\textwidth}
%                \centering
%                \includegraphics[width=0.999\textwidth]{plots/Chapter5/channel_vs_channel_differential/ref_channel69138.pdf}
%                \subcaption{Silicon channel reference.
%                }
%                \label{fig:tReso_vsCell_Siref}
%        \end{subfigure}
%        \hfill
%        \caption{Time resolution as a function of the cell energy in data, when
%          the calibrated ToA is compared to (a) the MCP measurement and (b) a reference silicon channel,
%          for pairs of cells on different modules (orange) and on the same module (blue).}
%        \label{fig:tReso_vsCell_MCP_Si}
%\end{figure}

The constant term of the black curve in \ref{fig:tReso_vsCell_MCP_HGC_ref} includes the contribution of \qty{\sim25}{\ps} due to the 
intrinsic timing resolution of the MCP system used as reference (\ref{fig:mcpreso}), providing a measurement
of the average per-cell asymptotic timing resolution of about \qty{80}{\ps}.\\
In \ref{fig:tReso_vsCell_Siref}, the difference between the constant terms determined from the same-module and different-module pairs
indicates strong correlations in the timing measurement of different channels within the same module.
To estimate this correlation, we fit to the data a model that assumes the same timing resolution for all silicon cells and one single correlation coefficient independent of the cell energy or its position within the module.
The fit yields a constant term for uncorrelated cells of about \qty{60}{\ps}
and a correlation coefficient $\rho\sim0.8$ between the timing measurements of cells in the same module.
This correlation was found to have a negligible impact on the results, as discussed in \ref{correlation}.
The difference between the per-cell constant terms when measured with the MCP and a with a silicon cell reference,
indicates the presence of an additional smearing of about \qty{50}{\ps} between the HGCAL prototype and the MCP system.
Although the source of this extra jitter could not be identified,
we believe that it is constant and random, and thereby does not affect the performance of the calibration procedure of \ref{sec:toa_recoCalib}.

To measure the intrinsic timing performance of the HGCAL prototype,
the calibrated TOA values are compared to an internal timing reference provided
by the average time of the shower measured with the calorimeter prototype, as described in \ref{sec:fullShower_perf}.
Such a quantity is independent of any offset between the HGCAL prototype and the MCP system and is dominated by the per-cell timing resolution.
The corresponding result is shown in \ref{fig:tReso_vsCell_MCP_HGC_ref} (blue squares) and fitted with the same resolution function.
The resulting energy-dependent term is essentially the same as that obtained when using the MCP as the reference, while the difference between the two constant terms is consistent with the intrinsic timing resolution of the MCP system plus the inferred extra global event jitter.
%
%\begin{figure}[hbtp]
%        \centering
%        \includegraphics[width=0.43\textwidth]{plots/Chapter5/timing_resolution_vs_cell_energy/sigmas_minus_shower_average_vs_cell_energy_electron.pdf}
%        \caption{Time resolution as a function of the cell energy in data, when
%          the calibrated ToA is compared to the average time of the shower measured with the HGCAL prototype.
%        }
%        \label{fig:tReso_vsCell_HGCref}
%\end{figure}

As a summary of \ref{fig:tReso_vsCell_MCP_Si}, the timing resolution representative of the average per-channel performance,
measured with the full readout chain, can be expressed as a function of the deposited energy as:
\begin{equation}
%        \sigma(\mbox{E}) \simeq \frac{13.46 \mbox{ ps}}{\mbox{E}}  \plus 61.72 \mbox{ ps}
        % \sigma(\mbox{E}) \simeq \sqrt{\frac{(\qty{13.46}{\ps}) ^2}{\mbox{E}^2} + (\qty{61.72}{\ps})^2 }
        \sigma^2(\mbox{E}) \approx \left(\frac{\qty{13.5}{\ns\cdot\MIP}}{\mbox{E}}\right)^2 + (\qty{62}{\ps})^2 
        \label{eq:toa_cell_perf}
%        \caption{\textbf{check numbers with plots}}
\end{equation}

This resolution agrees with the electronics specifications of the SKIROC2-CMS ASIC and is used for the smearing of the \GEANTfour-simulated hit timestamps for the analysis presented in the following sections.

%The asymptotic resolution measured on the time difference between each silicon cell and the MCP reference is
%shown in \ref{fig:tReso_Cells_summary} for the full set of calibrated cells. The resulting distribution has
%a rather Gaussian shape with a central value fully compatible with the constant term fit in \ref{fig:tReso_vsCell_MCPref}.

%\begin{figure}[hbtp]
%        \centering
%        \includegraphics[width=0.45\textwidth]{plots/Chapter5/channel_vs_MCP_constant_terms/channels_vs_MCP_constant_terms.pdf}
%        \caption{Constant term for single channels over the 136 calibrated cells.}
%        \label{fig:tReso_Cells_summary}
%\end{figure}

%\section{Timing performance of full showers and comparison with \GEANTfour-based simulation}

\subsection{Full shower performance}
\label{sec:fullShower_perf}

%Content: presenting the smearing model injected in the simulation [the extra jitter 
%has already been introduced and can be included in the model from the 
%               beginning]
%\textcolor{red}{Responsible: Arabella, other contributors: Axel, Shilpi, Thorben}

The timing performance measured for full showers in data is compared to the \GEANTfour simulation introduced in \ref{sec:datasets}.
Realistic timing values are simulated by smearing the TOA values from \GEANTfour with the average per-cell time resolution as discussed in \ref{sec:singleChannel_perf} and given in \ref{eq:toa_cell_perf}, including a term dependent on the energy deposited in the cell that is uncorrelated among the cells, and a constant term of about \qty{60}{\ps}. This constant term includes a contribution from the MCP measurement of
\qty{25}{\ps} and an additional jitter, discussed in \ref{sec:singleChannel_perf}, of about \qty{50}{\ps}. Both contributions to the constant term are correlated over all the cells.

%Content: present the procedure to compute the average time for the       
%               shower, as well as the fitted and expected sigmas, and present 
%               the corresponding  results for data and MC.
%               Average time of hits vs event [to confirm that the MCP-HGC bias
%               is event dependent, and the same for the 2 halves of the 
%               detector]

The average time of a shower, $\bar{t}$, is estimated as the weighted average over the times, $t_i$, of the ($n$) contributing cells:
\begin{equation}
  \bar{t} = \frac{\Sigma^{n}_{i = 1}{w_i t_i}}{\Sigma^{n}_{i = 1}{w_i}} \mbox{, where } w_i = \frac{1}{\sigma^2\left(E_{hit}\right)}
  \label{eq:avgT_shower}
\end{equation}
These shower time values are then fitted with a Gaussian function in bins of the particle beam energy.
This allows to extract the average time and its standard deviation as a function of the impinging particle energy. The standard deviation from this fit is referred to in the following as the \emph{fitted resolution}.
% To consolidate the characterization of the dataset
For comparison, the average of the per-event uncertainty is also computed, and referred to as \emph{expected resolution}, $\sigma_{\bar{t}}$, in the following:
\begin{equation}
  \sigma_{\bar{t}} = \frac{1}{\sqrt{\Sigma^{n}_{i = 1}{w_i}}}
  \label{eq:expS_shower}
\end{equation}

\begin{figure}[hbtp]
%   \captionsetup[subfigure]{aboveskip=-1pt,belowskip=-1pt}
%   \centering
  \hfill
  \begin{subfigure}[b]{0.42\textwidth}
    % \centering
    % \hfill
    \includegraphics[width=0.999\textwidth]{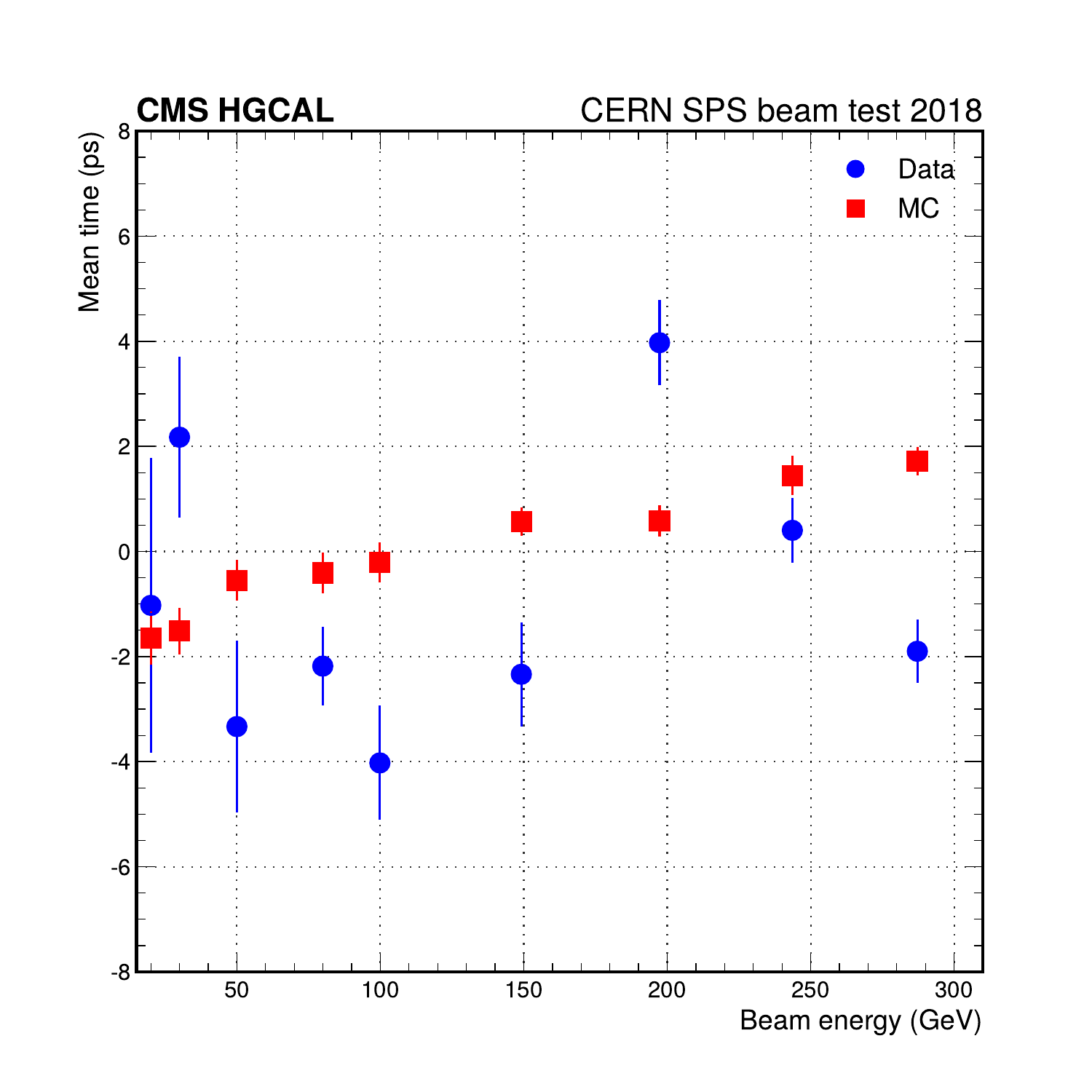}
    \subcaption{Average shower time showing an accuracy better than \qty{10}{\ps} across the full range of energies probed.}
    \label{plot:avgT_vsE}
  \end{subfigure}
  \hfill
  \begin{subfigure}[b]{0.42\textwidth}
    % \centering
    \includegraphics[width=0.999\textwidth]{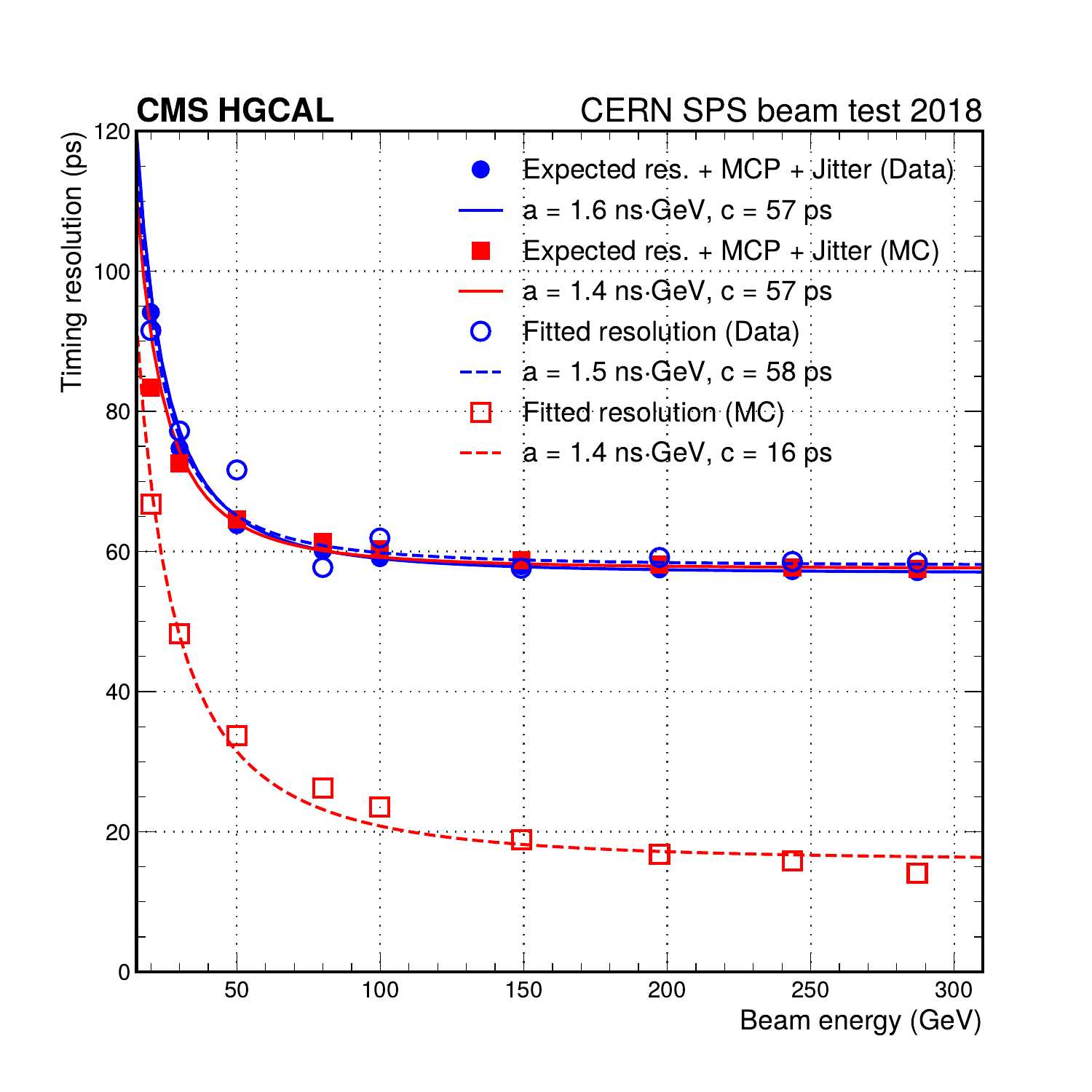}
    \subcaption{Fitted and expected shower timing resolutions. The fitted resolution in data is dominated by the additional jitter between the calorimeter prototype and the MCP detectors.}
    \label{plot:reso_vsE}
  \end{subfigure}
    \hfill
  \caption{For positron showers contained in the HGCAL prototype, their (a) average time and (b) timing resolution as a function of the beam energy, for both beam data and simulated Monte Carlo (MC) data.
  For beam data, both the fitted and the expected resolutions are shown, as discussed in the text.
  \label{fig:avgT_reso_dataMC}
  }
\end{figure}

\ref{fig:avgT_reso_dataMC} shows the results for the average shower time and the fitted and expected resolutions, as well as a good agreement between data and simulation,
when including all previously-discussed smearing terms.
The presence of the additional jitter between the calorimeter prototype and the MCP detectors clearly deteriorates the observed timing resolution performance for full showers.
The constant term of about \qty{58}{\ps} is compatible with the sum in quadrature of
the MCP intrinsic resolution of about \qty{25}{\ps} (\ref{fig:mcpreso}),
the expected calorimeter performance of about \qty{16}{\ps} (\ref{plot:reso_vsE}),
and an additional jitter of about \qty{50}{\ps}.

\subsubsection{Intrinsic timing resolution of the HGCAL prototype}%Difference between even and odd layers}
%Content: from this we can extrapolate what the HGC-prototype standalone 
%               timing performance can be (in case of negligible extra 
%               contribution and using all the layers)

The intrinsic timing performance of the HGCAL prototype can be characterized
in spite of the experimental jitter observed between the calorimeter and the MCP devices.
This is achieved by splitting the calorimeter into two equivalent halves for analysis purposes, each half acting as a reference and a target timing measurement, respectively.
Even and odd layers are considered separately, and the time difference between the two \emph{half-showers} reconstructed in each of the halves of the calorimeter is taken on an event-by-event basis.
The intrinsic time resolution is then determined under the assumption of similar time resolution of both halves, by taking the standard deviation of the time difference between the halves divided by $\sqrt{2}$.
%The resolution of such a time difference when divided by $\sqrt{2}$ is equivalent to the time measurement provided by a detector of similar depth and all the sampling layers.

\begin{figure}[hbtp]
        \centering
        \includegraphics[width=0.45\textwidth]{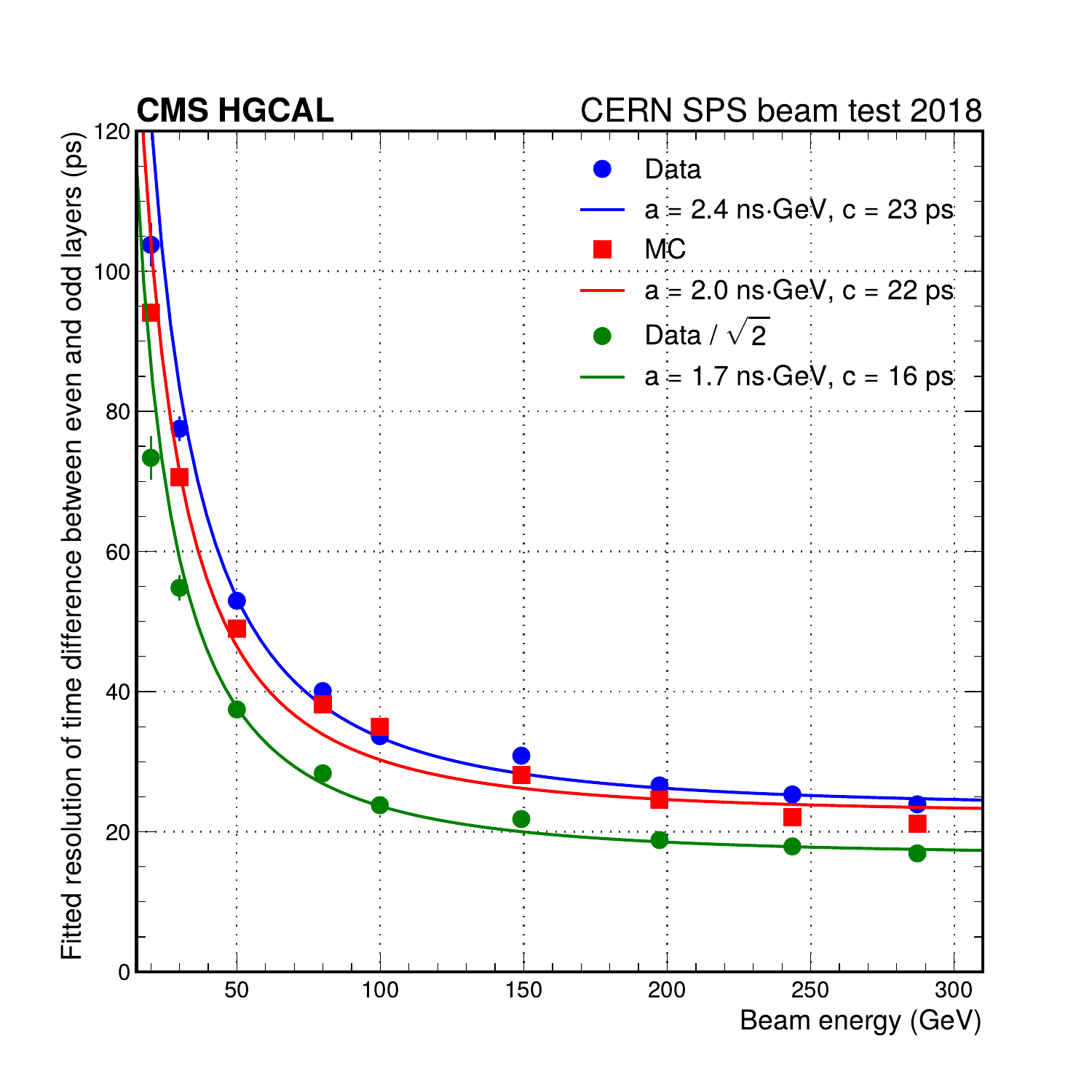}
        \caption{Resolution of the event-by-event difference between the time of positron half-showers
          computed with only even layers and only odd layers separately, as obtained for beam data (blue) and simulated Monte Carlo (MC) data (red).
          The green markers correspond to beam data resolutions divided by $\sqrt{2}$, and are an estimate of the
          performance expected if all the layers were used for the shower time estimation.
          These measurements are not affected by the jitter between MCP and HGCAL, as discussed in the text.
        \label{fig:reso_EO}
        }
\end{figure}

\ref{fig:reso_EO} shows the half-shower timing resolution as a function of the beam energy.
The estimated value of the energy-dependent term, $a$, is a factor of $\sqrt{2}$ larger than what is found in \ref{plot:reso_vsE}, which is consistent with the using half the number of sampling layers.
The constant term, on the other hand, is almost twice smaller than when using the MCP as reference, due to the absence of the jitter between the MCP and the HGCAL prototype and the green markers in \ref{fig:reso_EO} illustrate the estimated performance when using all layers for the shower timing determination.

The constant term of about \qty{16}{\ps} for the full calorimeter prototype estimates in \ref{fig:reso_EO} is consistent with the constant term for simulated data that does not include the \qty{50}{\ps} correlated jitter that is shown in \ref{plot:reso_vsE}.

\subsubsection{Timing properties of the shower}
%Content: Showing the scaling of performance vs beam energy and beam 
%              profile, and if available also present the average time vs distance 
%             from shower axis and the time development of shower

\begin{figure}[hbtp]
        % \captionsetup[subfigure]{aboveskip=-1pt,belowskip=-1pt}
        % \centering
        \hfill
        \begin{subfigure}[b]{0.45\textwidth}
                \centering
                \includegraphics[width=0.999\textwidth]{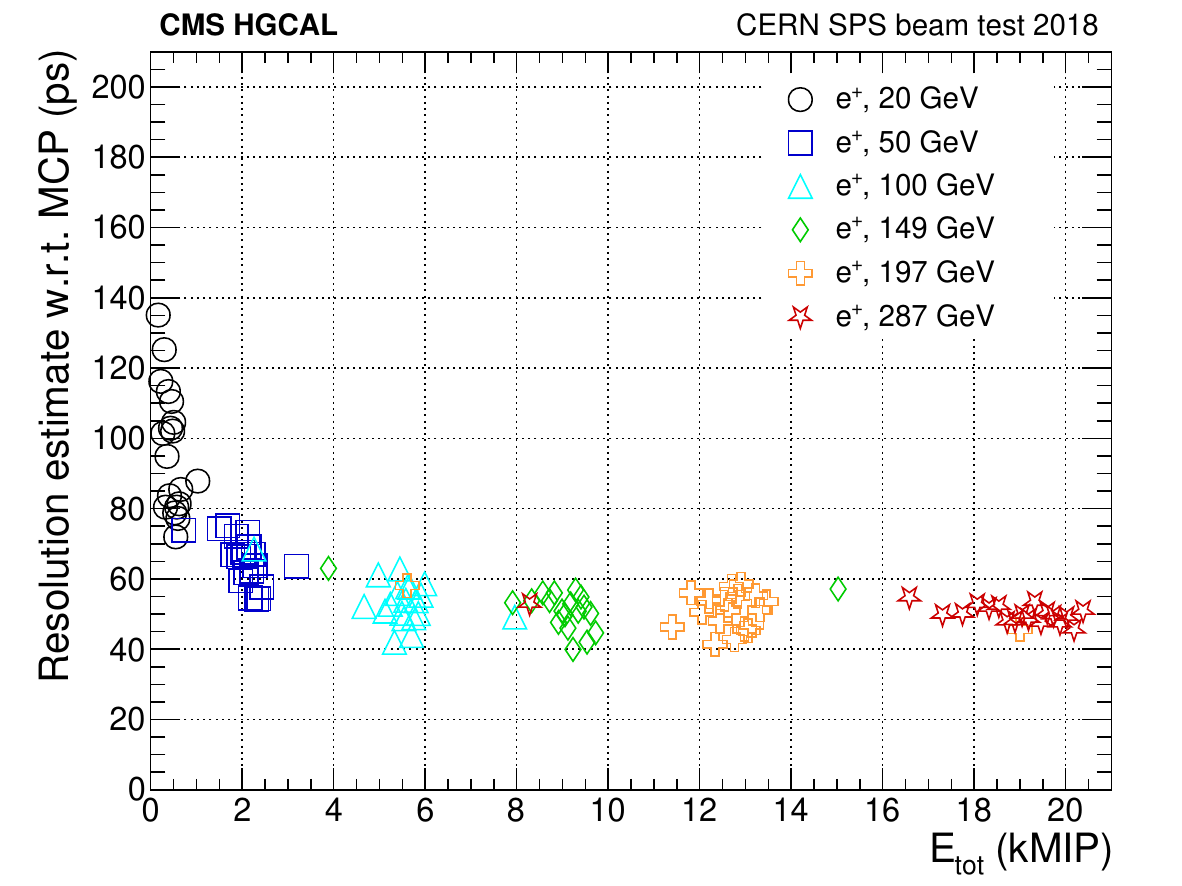}
                \subcaption{}
                % Resolution dependency on energy sum.
                % }
                \label{plot:reso_Esum}
        \end{subfigure}
        \hfill
        \begin{subfigure}[b]{0.45\textwidth}
                \centering
                \includegraphics[width=0.999\textwidth]{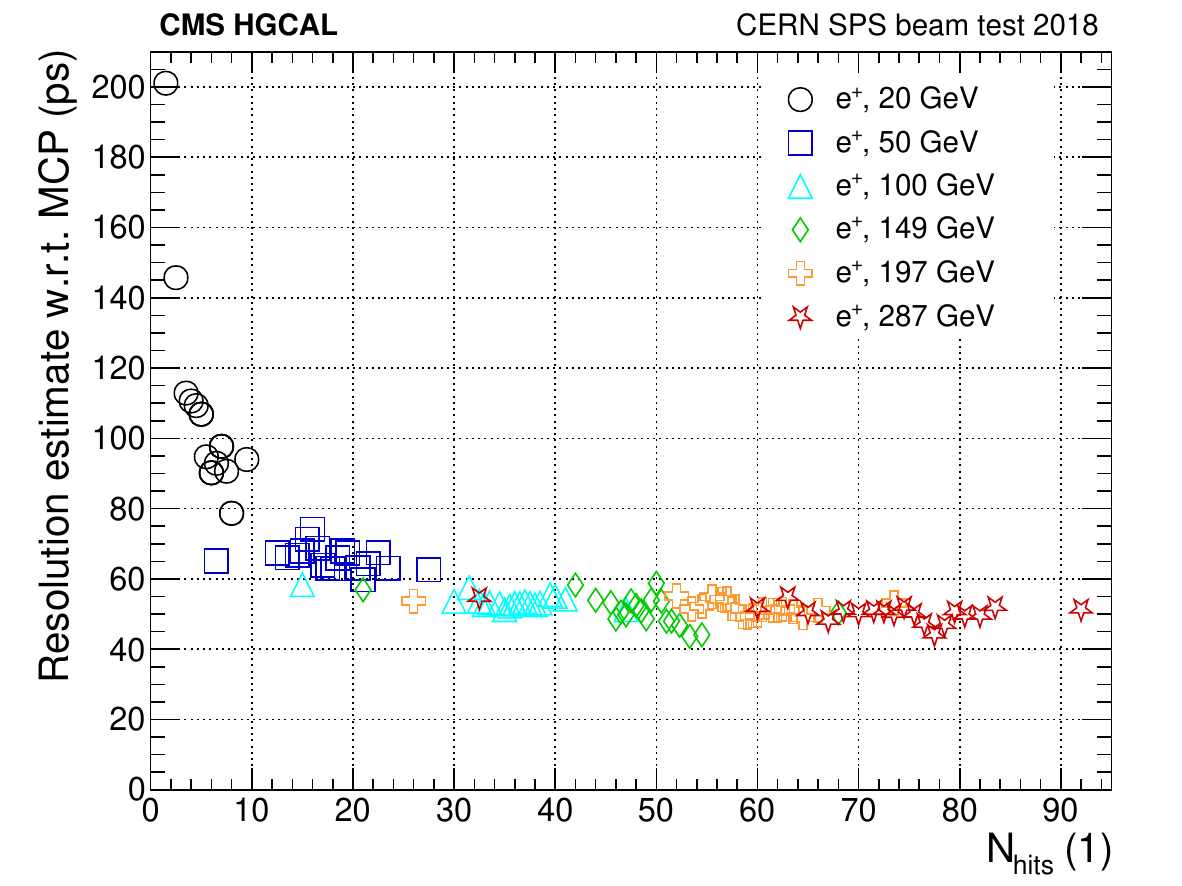}
                \subcaption{}
                %   Resolution dependency on total number of hits.
                % }
                \label{plot:reso_Nhits}
        \end{subfigure}
        %\hfill
        %\begin{subfigure}[b]{0.42\textwidth}
        %        \centering
        %        \includegraphics[width=0.999\textwidth]{plots/Chapter6/differential_resolution/esum_pions.pdf}
        %        \subcaption{
        %        }
        %        \label{plot:reso_Esum_pion}
        %\end{subfigure}
        %\hfill
        %\begin{subfigure}[b]{0.42\textwidth}
        %        \centering
        %        \includegraphics[width=0.999\textwidth]{plots/Chapter6/differential_resolution/nhits_pions.pdf}
        %        \subcaption{
        %        }
        %        \label{plot:reso_Nhits_pion}
        %\end{subfigure}
        \hfill
        \caption{Time resolution of the reconstructed positron showers as a function of (a) the energy sum and of (b) the number of hits. The smooth variation of the resolution with these two quantities shows that the shower time determination is stable with respect to varying beam conditions, including different beam profiles.
        \label{figure:differentialReso}
        }
\end{figure}

Since the timing resolution of a shower depends both on the energy of the hits and on the number of hits, the time resolution is studied as a function of these two quantities.
The results for beam data are shown in \ref{figure:differentialReso} for reconstructed showers using all layers.
The resolution is observed to scale according to the expectation, with a continuous trend across different beam particle energies and hit multiplicities. %, that also have different beam profiles.
% This agreement shows that the shower time determination is stable with respect to varying beam conditions, including different beam profiles.
The smooth trends observed in \ref{figure:differentialReso} are all the more relevant when considering the underlying variety of beam conditions, such as the beam profile that changes substantially with the beam energy.
%The differential measurements obtained with pion data again show that only a part of the full shower is probed,
%and the few points corresponding to large values of the x-axis are affected by important statistical fluctuations.\\

%\begin{figure}[hbtp]
%  \centering
%  \includegraphics[width=0.45\textwidth]{plots/Chapter6/avgT_vsRadialdistance.pdf}
%  \label{plot:avgT_vsRadiald}
%  \caption{
%  }
%\end{figure}

\begin{figure}[hbtp]
        \captionsetup[subfigure]{aboveskip=-1pt,belowskip=-1pt}
        \centering
        \hfill
        \begin{subfigure}[b]{0.49\textwidth}
                \centering
                \includegraphics[width=\textwidth]{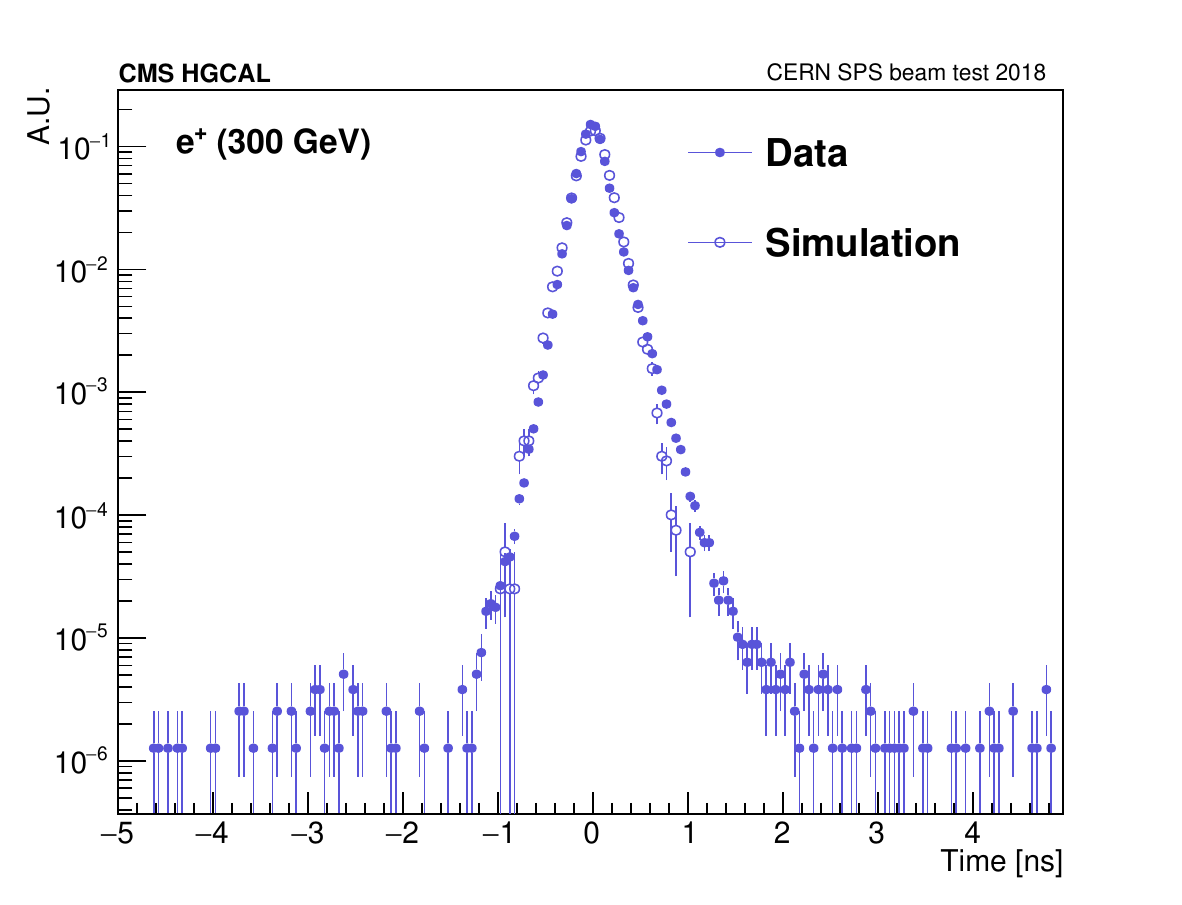}
                \subcaption{}
                % Time distribution of hits.
                % }
                \label{plot:hitsT_inShower}
        \end{subfigure}
        \hfill
        \begin{subfigure}[b]{0.49\textwidth}
                \centering
                \includegraphics[width=\textwidth]{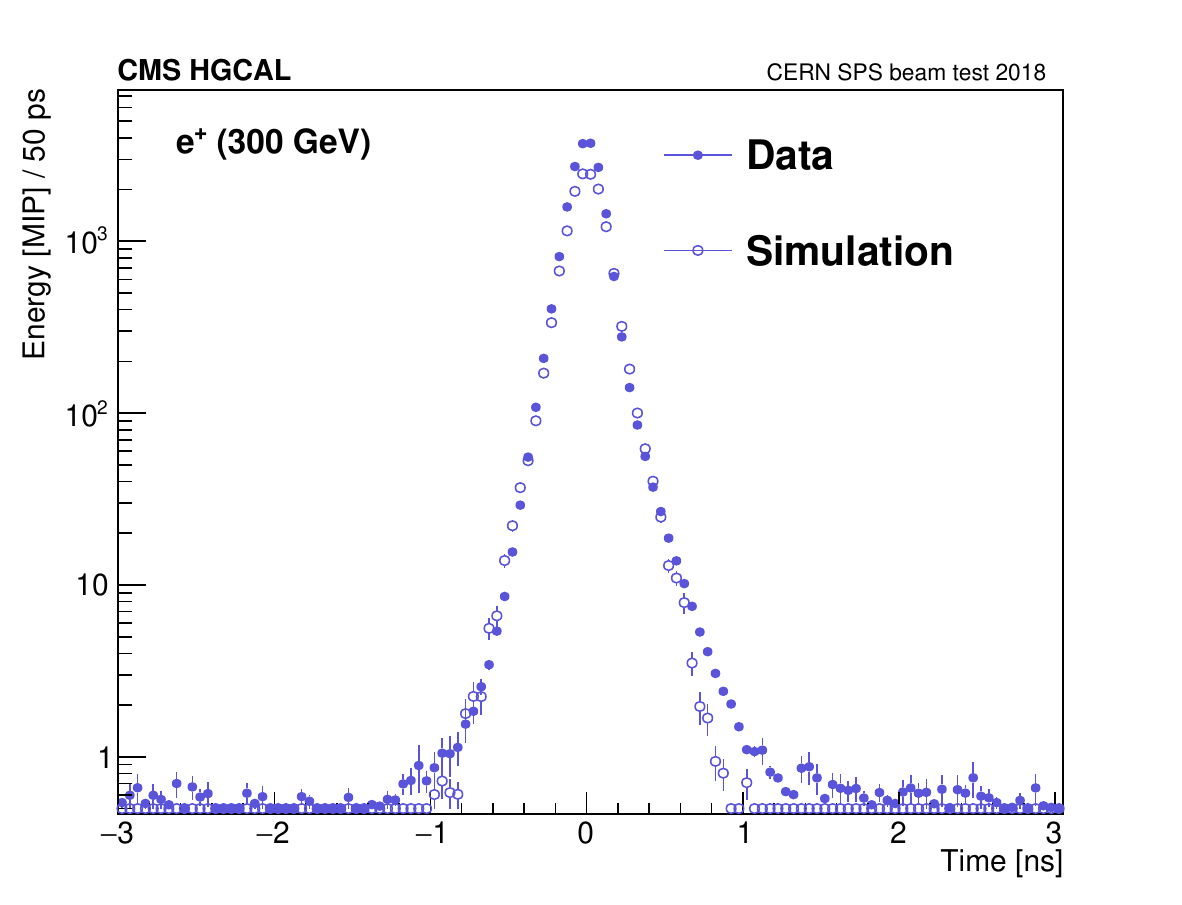}
                \subcaption{}
                % Time distribution of hit energy.
                % }
                \label{plot:ene_wrtT_inShower}
        \end{subfigure}
        \hfill
        \caption{For the hits used in the reconstruction of the same \qty{300}{\GeV} positron showers,
          (a) time distribution of the fraction of hits and (b) the average and standard deviation of the energy of hits
          as a function of their calibrated time. Beam data in full markers and simulated data in empty markers.
        }
\end{figure}

%\noindent 
The time distribution of the fraction of hits that were calibrated and used in the reconstruction of \qty{300}{\GeV} positron showers is displayed in \ref{plot:hitsT_inShower} for both beam data and simulated data, showing a good agreement between the two.
For the same showers, \ref{plot:ene_wrtT_inShower} shows the energy distribution average and standard deviation of the
hits as a function of their calibrated time. One can see that the most energetic component of the shower is deposited at times around zero by construction of the calibration. Also in this case a reasonable agreement is found between data and simulation.

\section{Discussion and conclusion}
\label{sec:concl}
%Content: - discussion on the origin of the extra-jitter
%               - performance of the HGC-prototype branch indicating that the clock tree 
%                 here is well under control
%                     - underline that these results represent the first timing measurement with 
%                  few tens of ps precision
%                    - outlook
%                      - making it clear that these results are not the final HGCAL ones, but this 
%               is the first HGC-timing prototype tested
%\textcolor{red}{Contributors: Arabella, Artur}

We presented the timing performance of the first HGCAL prototype for positron showers.
The focus of the analysis was to characterize the timing performance of single channels,
%in comparison with the designers' specifications,
perform measurement with full showers, and compare the results to \GEANTfour simulation.

After the multi-step calibration of the TOA response, \num{116} readout channels in the central electromagnetic section could be fully calibrated, with an average asymptotic per-channel timing resolution of about \qty{60}{\ps}, consistent with the electronics specifications.
The time measurement provided by the MCP system was exploited as a reference throughout the calibration process.
The MCP detector itself was measured to have a time resolution of the order of \qty{25}{\ps} for the average energy of selected positrons.
An additional jitter of about \qty{50}{\ps} between the MCP and HGCAL systems was found, and its origin could not be identified. This jitter is assumed to be constant and random, such that its presence does not compromise the calibration.
% Due to limited amounts of data, only channels in the central electromagnetic section of the HGCAL prototype were calibrated.
The timing of full positron showers was measured and compared with a simulation where the ideal timing information was smeared according to the single-channel resolution model derived from data.

The intrinsic performance of the HGCAL setup was tested by splitting the calorimeter prototype in two equivalent halves and taking the time difference between the two halves when reconstructing the same shower.
The measured time resolution was found to be in agreement with simulation.
% results and had a constant term of about \qty{23}{\ps} that corresponded to the time performance of a detector with equivalent total depth and only half the layers.

\begin{figure}[hbtp]
	%\captionsetup[subfigure]{aboveskip=-1pt,belowskip=-1pt}
	\centering
	\includegraphics[width=0.6\textwidth]{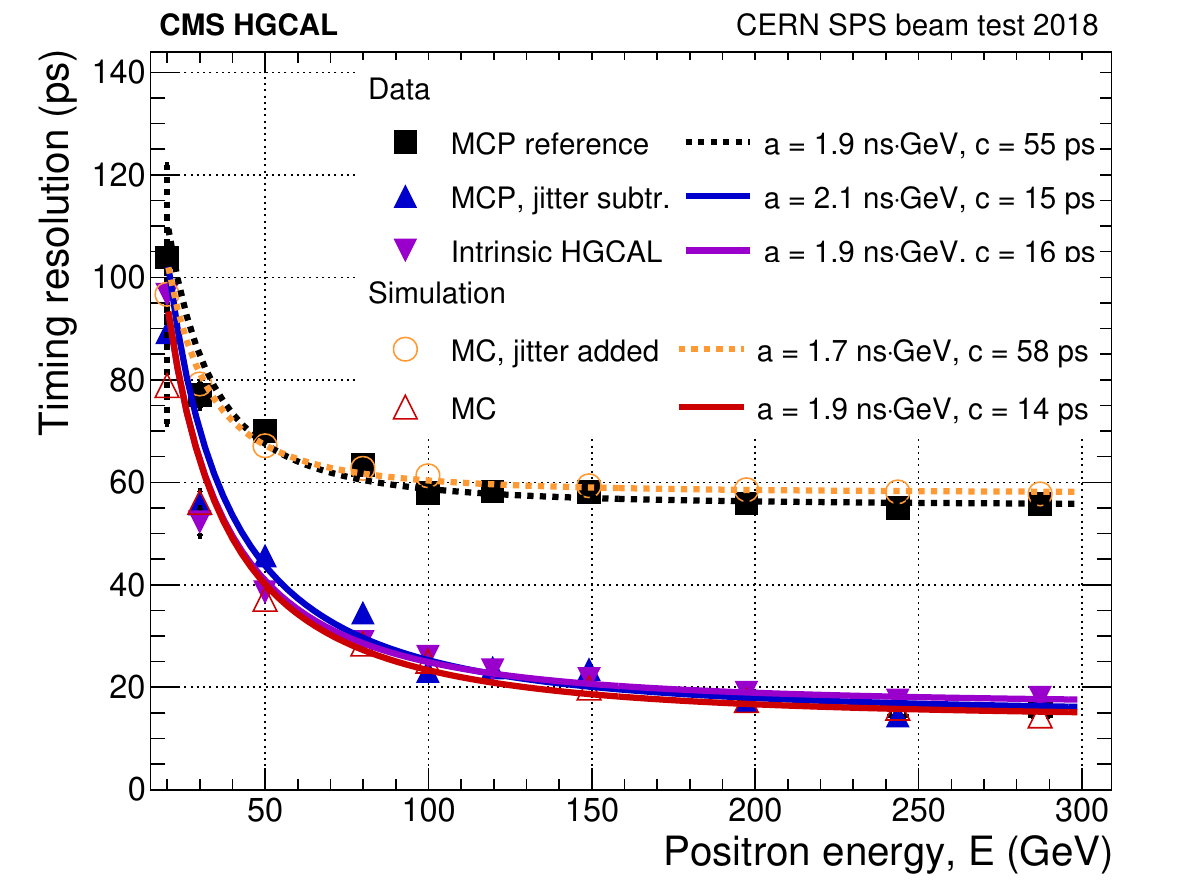}
	\caption{Comparison of the HGCAL prototype timing resolution for positron showers in data and simulated samples. For data, the resolution is measured for all layers using the MCP as a reference (black squares) as well as using only half the layers with respect to the other half and assuming they have identical resolution (purple triangles).
	Other measurements in the figure allow to cross-check and confirm the hypothesis that a global jitter between the MCP and HGCAL systems was present in the data.
	}
	\label{plot:resolution_conslusion}
\end{figure}

\ref{plot:resolution_conslusion} summarizes the measured resolution for positron showers, showing good agreement after taking into consideration the observed jitter between the MCP and HGCAL systems.
% The results also show the intrinsic HGCAL prototype resolution having a constant term distinctly under \qty{20}{\ps}.

This work represents the first measurement of HGCAL timing performance with a precision of tens of picoseconds. It also demonstrates the stability of the clock distribution used in this prototype.
The results can be understood as experimental evidence of the possibility to achieve $\mathcal{O}(\qty{10}{\ps})$ timing resolutions with the new CMS high-granularity endcap calorimeter. 
This  timing performance is expected to enable effective separation of pile-up interactions and, with it, contribute towards a successful operation of the CMS detector at the HL-LHC.

\acknowledgments

% Based off of https://arxiv.org/abs/2211.04740
We thank the technical and administrative staffs at CERN and at other CMS institutes for their
contributions to the success of the CMS upgrade program. We acknowledge the enduring support
provided by the following funding agencies and laboratories: BMBWF and FWF (Austria); CERN;
CAS, MoST, and NSFC (China); MSES and CSF (Croatia); CEA, CNRS/IN2P3 and P2IO LabEx
(ANR-10-LABX-0038) (France); SRNSF (Georgia); BMBF, DFG, and HGF (Germany); GSRT
(Greece); DAE and DST (India); MES (Latvia); MOE and UM (Malaysia); MOS (Montenegro);
PAEC (Pakistan); FCT (Portugal); JINR (Dubna); MON, RosAtom, RAS, RFBR, and NRC KI
(Russia); MoST (Taipei); ThEP Center, IPST, STAR, and NSTDA (Thailand); TUBITAK and
TENMAK (Turkey); STFC (United Kingdom); and DOE (USA).

\appendix
\section{Example of hit timestamp calibration constants}
\label{appendix:calib_constants}

\begin{table}[h]
	\caption{Example of fitted calibration constants from \ref{eq:toa_tw_param,eq:toa_twcalib,eq:toa_twmodule}. The corresponding functions are plotted in \ref{plot:TOA_calib_rise,plot:TW_calib_rise,plot:ESUMTW_calib_rise}.
    Values are rounded to their least significant digits.}	
	\centering
	\begin{tabular}{cl||cl||cl}
		\textbf{Parameter} & \textbf{Value} & \textbf{Parameter} & \textbf{Value} & \textbf{Parameter} & \textbf{Value} \\
		$\Theta_{1, 1}^\text{TOA}$ & \qty{-14.21}{\ns} & $\Theta_{1, 1}^\text{TW}$ & \qty{2.97}{\ps\per\MIP} & $p_0$ & \qty{314}{\ps}\\
		$\Theta_{1, 2}^\text{TOA}$ & \qty{30.70}{\ns} & $\Theta_{1, 2}^\text{TW}$ & \qty{-0.73}{\ns} & $p_1$ & \qty{-0.24}{\ps\per\MIP}\\
		$\Theta_{1, 3}^\text{TOA}$ & \qty{3.62}{\ns} & $\Theta_{1, 3}^\text{TW}$ & \qty{-177}{\ns\MIP} & $p_2$ & \qty{-2e-5}{\ps\per\square\MIP} \\
		$\Theta_{1, 4}^\text{TOA}$ & \num{1.253} & $\Theta_{1, 4}^\text{TW}$ & \qty{-0.8}{\MIP} & $p_3$ & \qty{4e-8}{\ps\per\cubic\MIP} \\
		$\Theta_{2, 1}^\text{TOA}$ & \qty{-10.00}{\ns} & $\Theta_{2, 1}^\text{TW}$ & \SI{0.05}{\ps\per\MIP} & $p_4$ & \qty{-1.2e-11}{\ps\per\MIP\tothe{4}} \\
		$\Theta_{2, 2}^\text{TOA}$ & $\hat{f}\left(0.65\middle|\vec{\Theta}_{1}^{TOA}\right)$ & $\Theta_{2, 2}^\text{TW}$ & $\hat{f}\left(E_\text{TOT}\middle|\vec{\Theta}_{1}^{TW}\right)$ & & \\
		$\Theta_{2, 3}^\text{TOA}$ & \qty{5.53}{\ns} & $\Theta_{2, 3}^\text{TW}$ & \qty{-730}{\ns\MIP} & & \\
		$\Theta_{2, 4}^\text{TOA}$ & \num{1.298} & $\Theta_{2, 4}^\text{TW}$ & \qty{-150}{\MIP} & & \\		
	\end{tabular}
	\label{table:calibconstants_fitted}
\end{table}
\section{Correlation effects}
\label{correlation}
To evaluate the impact of the large in-module timing correlation discussed in \ref{sec:singleChannel_perf}, the full shower performance was
re-evaluated by replacing \ref{eq:avgT_shower,eq:expS_shower} with the more general:
\begin{equation}
  \bar{t} = \sigma^2_{\bar{t}} \left( J^TWX \right) \mbox{   and   }  \sigma_{\bar{t}}^2 = \left( J^TWJ \right)^{-1},
  \label{eq:correlation_shower}
\end{equation}
where $X = [t_1,..., t_n]$, $J = [1,..., 1]^T$, and $W = C^{-1}$, where $C$ is the covariance matrix among the $t_{i}$ measurements:
\begin{equation}
      C = \begin{pmatrix*}[c]
        \sigma_1^2    &  \ldots   & \sigma_{ij}   \\
        \vdots        &  \ddots   & \vdots          \\    
        \sigma_{ij}   &  \ldots   & \sigma_n^2   \\     
      \end{pmatrix*}
  \text{  , where  } \sigma_{ij} = \rho \cdot \sigma_i \sigma_j = 0.8 \cdot \sigma_i^2 \text{  for off-diagonal terms.}
  \label{eq:covMatrix}
\end{equation}
The analysis was repeated following the same procedures and the obtained results differ from those shown in \ref{fig:avgT_reso_dataMC} by only a few picoseconds.

The observed large correlation has a small impact
on the quoted performance, and the remainder of the results reported in this paper does not include the
correlation model discussed in this section.

\bibliographystyle{JHEP}
\bibliography{bib/bib}

\end{document}